\documentclass[aps,superscriptaddress,showpacs,nofootinbib,eqsecnum,prd]{revtex4}

\usepackage[bf,SL,BF]{subfigure} 
\usepackage{verbatim} 
\usepackage{epsfig}  
\usepackage{amsmath}
\usepackage{amssymb}
\usepackage{textcomp}
\usepackage{amssymb}
\input{epsf}
\def\be{\begin{equation}} 
\def\ee{\end{equation}} 
\def\bea{\begin{eqnarray}}
\def\eea{\end{eqnarray}}

\def\beq{\begin{equation}}
\def\eeq{\end{equation}}
\def\beqa{\begin{eqnarray}}
\def\eeqa{\end{eqnarray}}

\newcommand{\bolsy}{\boldsymbol}

\begin{document}

\title{Dynamical Generation of a Repulsive Vector Contribution to the Quark Pressure  }

 \author{Tulio E. Restrepo}  
\affiliation{Departamento de F\'{\i}sica, Universidade Federal de Santa
  Catarina, 88040-900 Florian\'{o}polis, Santa Catarina, Brazil}  
  
\author{Juan Camilo Macias}  
\affiliation{Departamento de F\'{\i}sica, Universidade Federal de Santa
  Catarina, 88040-900 Florian\'{o}polis, Santa Catarina, Brazil}    

\author{Marcus Benghi Pinto}  
 \email{marcus.benghi@ufsc.br} 
\affiliation{Departamento de F\'{\i}sica, Universidade Federal de Santa
  Catarina, 88040-900 Florian\'{o}polis, Santa Catarina, Brazil}  
  
 \author{Gabriel N. Ferrari} 
\affiliation{Departamento de F\'{\i}sica, Universidade Federal de Santa
  Catarina, 88040-900 Florian\'{o}polis, Santa Catarina, Brazil}

\begin{abstract}
Lattice  QCD results  for the coefficient $c_2$ appearing in the Taylor expansion of the pressure show that  this quantity raises with the temperature towards the Stefan-Boltzmann limit. On the other hand, model approximations predict that when a vector repulsion, parametrized by $G_V$, is present this coefficient reaches a maximum  just after $T_c$ and then deviates from the lattice predictions. Recently, this discrepancy has been  used as a guide to constrain the (presently unknown) value of $G_V$ within the framework of effective models at large-$N_c$ (LN).  In the present investigation we show that, due to finite $N_c$ effects, $c_2$ may also develop a maximum even when $G_V=0$   since a vector repulsive term can be dynamically  generated by exchange type of radiative corrections. Here we apply the the Optimized Perturbation Theory (OPT) method to the two flavor Polyakov--Nambu--Jona-Lasinio model (at $G_V=0$) and compare the results with  those furnished by lattice simulations an by the LN approximation at $G_V=0$ and also at $G_V \ne 0$. The OPT numerical results  for $c_2$ are impressively  accurate for $T \lesssim 1.2\, T_c$ but, as expected, predict that  this quantity develops a maximum at high-$T$. After identifying the mathematical origin  of this extremum  we argue  that such a  discrepant behavior may naturally arise within these effective quark models (at $G_V=0$)  whenever the first $1/N_c$ corrections are taken into account. We then interpret this hypothesis  as an indication that beyond the large-$N_c$ limit the correct high temperature (perturbative) behavior of $c_2$ will be faithfully described  by effective models only if they also mimic the asymptotic freedom phenomenon.

\end{abstract}


\maketitle

\section{Introduction}

It is generally believed   that effective theories used to describe  compressed strongly interacting matter should include  vector channels \cite{buballa, bonanno,lenzi2012,logoteta,shao2013,sasaki2013,masuda2013,baym}  such as the ones which appear in the Walecka model for nuclear matter \cite {walecka}  and   in the extended version of the Nambu--Jona-Lasinio model (NJL) for quark matter  \cite {volker}. To emphasize its importance let us point out few recent applications which consider this channel in the framework of the NJL model starting with Ref. \cite {gvstars} where  the three flavor version of this theory has been  used to reproduce the equation of state (EoS) for  cold magnetized quark matter. In agreement with Ref. \cite {robson} the results show that the magnetic field and the vector channel tend to influence the first order chiral transition in opposite ways: while the first softens the EoS the second  hardens it so that higher stellar masses may be reproduced giving further insight to the modeling of stellar objects such as    the two recently measured pulsars, 
PSR J1614-2230 \cite{Demorest10} and   PSR J0348+0432 \cite{j0348}, whose masses are about $2M_\odot$. Another timely important application \cite {ko} shows that the presence of a vector interaction   is crucial for the NJL theory to reproduce the measured relative elliptic flow differences between nucleons and 
anti-nucleons as well as between kaons and antikaons at energies 
carried out in the Beam-Energy Scan program of the Relativistic Heavy Ion Collider (RHIC). Also, in Ref. \cite {volker1}, 
it has been proposed  that the large elliptic flow at RHIC could be described by single-particle dynamics with
a repulsive interaction.  As a final example, let us recall that  although most of  investigations  seem to support the QCD critical point (CP), an interesting observation against its existence has been advanced  by de Forcrand  and Philipsen \cite {deForcrand}.  A possible explanation for this disagreement has been given in Ref.  \cite {Fukushima} where it was suggested   that a  strong (repulsive) vector coupling could conciliate the results found in Ref. \cite {deForcrand} with the existence of a CP in the QCD phase diagram.
In practice, within the NJL model, a vector channel can be  easily implemented by adding a term such as $- G_V ({\bar \psi} \gamma^\mu \psi)^{2}$ to the original Lagrangian density \cite {buballa,volker}. Then, within the large-$N_c$ approximation (LN) only the zeroth component survives so that  the {\it net} effect produced by this channel is to add a term like $-G_V \rho_q^2$ to the pressure (where $\rho_q$ represents the LN quark number density)  weakening (strengthening) the first order transition when $G_V$ is positive (negative) \cite {Fukushima2}. As a result, in the repulsive case ($G_V>0$),  the first order transition region covers a smaller range of temperatures as compared to the $G_V=0$ case while the coexistence chemical potential for a given temperature is shifted to a higher value. Then, as a consequence, the CP happens at smaller temperatures and higher chemical potentials than in the case of vanishing $G_V$.  Despite its importance, fixing $G_V$ in a non renormalizable model such as the NJL is a  delicate task. The reason is that the divergent integrals appearing in typical NJL evaluations are usually regulated by a sharp ultra violet momentum cut-off, $\Lambda$, which cannot be removed by a systematic redefinition of the {\it original} parameters as in a renormalizable theory. To deal with this situation one  considers  $\Lambda$ to be a new ``parameter" which sets the maximum energy scale at which the model predictions can be trusted. Then, the original parameters together with $\Lambda$ are  fixed by requiring the model to reproduce the phenomenological values of physical observables. For example, in the standard two flavor version of the NJL model the scalar-pseudoscalar coupling ($G_S$), the current quark mass ($m_c$) and $\Lambda$ are adjusted so as to reproduce the pion mass  ($m_\pi \simeq 135 \, {\rm MeV}$), the pion decay constant ($f_\pi \simeq 93\, {\rm MeV}$) and the quark condensate ($\langle {\bar \psi} \psi \rangle^{1/3} \simeq 250 \, {\rm MeV}$) which yields $\Lambda \sim 560-670 \, {\rm MeV}$,  $G \Lambda^2 \sim 2-3.2$ , and $m_c \sim 5 - 5.6\, {\rm MeV}$ (see Ref. \cite {buballa} for a complete discussion). However,  fixing $G_V$ poses and additional problem since this quantity should be fixed using the $\rho$ meson mass which, in general, happens to be higher than the maximum energy scale set by $\Lambda$. At present,  the vector term  coupling  $G_V$ cannot be determined from experiments and lattice QCD simulations (LQCD) but  eventually, the combination of neutron star observations and the energy scan  of the  phase-transition signals 
at FAIR/NICA may provide us some hints on its precise numerical value.  While many authors  consider $G_V$ to be a  free parameter, whose value ranges between $0.25 \, G_S$ and $0.5 \, G_S$ \cite {gv1,gv2}, others try to fix it in different ways as in Refs.   \cite{klimt, hanauske, sakai, kashiwa, odilon,hatsuda} predicting $0.3 \le G_V/G_S \le 3.2$ so that the true value remains undetermined.

 At this point one should note that, due to the Fierz transformations, when going beyond the LN, or mean field, level one may induce radiative (exchange like) corrections which produce similar physical effects to those caused by a classical (tree) term such as $-G_V({\bar \psi} \gamma^\mu \psi)^{2}$ \cite {klevansky}. This is precisely what has been observed in an application of the nonperturbative Optimized Perturbation Theory (OPT)
method to the two flavor NJL model  with vanishing $G_V$ \cite {prc}.  The OPT results for the NJL  phase diagram  show that  the $1/N_c$ radiative corrections induced by this approximation reproduce the same qualitative features (weakening of the first order chiral transition) obtained by considering the model at large-$N_c$ with an explicit repulsive vector channel. This is because the OPT two loop contributions add a term like $-G_S \rho_q^2/(N_c N_f)$ to the pressure (recall that the LN net contribution goes as $-G_V \rho_q^2$). In Ref. \cite {prc1} the OPT (with $G_V=0$) was shown to produce results which are qualitatively similar to those obtained in Ref. \cite {Fukushima} with the LN approximation at $G_V \ne 0$.
This relationship between the OPT, at $G_V=0$, and the large-$N_c$ approximation, at $G_V \ne 0$, has been recently investigated in great detail in the framework of the abelian NJL at finite densities and zero temperature in Ref. \cite {ijmpe}. 
The most obvious advantage this alternative technique offers  with regard to  dense quark matter evaluations is that a more realistic description  can be obtained without the need to explicitly include the, so far, undetermined $G_V$ parameter.  Eventually, the same type of results could be achieved by going beyond the LN (one loop) level but, in practice, incorporating finite $N_c$ corrections  in a typical $1/N_c$ expansion is not always an easy task  since an infinite series of contributions has to be resummed \cite {root}. On the other hand, by combining {\it perturbative} evaluations with a variational optimization procedure the OPT offers a nonperturbative alternative to go beyond the large-$N_c$ limit. Having such an alternative  can be particularly useful in the analysis of compressed quark matter since, due to sign problem, QCD is  not yet completely accessible to lattice simulations when $\mu \ne 0$.

The OPT has already established
itself as a powerful method in dealing with  critical theories as the Bose-Einstein condensation where this method and its
different variations have provided some of the most precise analytical
results regarding the shift in the critical temperature for weakly
interacting homogeneous Bose gases~\cite{bec}. Other applications to
condensed matter situations include a precise evaluation of the
critical density for polyacetylene \cite{poly}.  Recently, Kneur and Neveu \cite{rgopt1} have improved the method with renormalization group properties  to evaluate $\Lambda_{\rm MS}^{\rm QCD}$~\cite{rgoptqcd1} and
$\alpha_S$~\cite{rgoptqcd2}.  The OPT was also instrumental in
the determination of the phase diagram of the massless GN model in 2+1
dimensions at finite $T$ and $\mu$ \cite {prdgn3d,plbgn3d}.   

Here, our first aim is to  extend the previous   OPT-NJL  applications \cite{prc,prc1}  to  the two flavor  Polyakov--NJL model (PNJL) which, by incorporating confinement, represents a more realistic theory.   
Technically, this extension is not completely straightforward  and for this reason we present details associated to the evaluation of color traces over two loop (exchange) contributions.  We then evaluate the PNJL free energy in order to obtain quantities such as the quark number density and the quark number susceptibility. Our numerical results are compared with the ones produced by the LN approximation, at $G_V=0$ and $G_V \ne 0$,  as well as with those produced by LQCD simulations. As we shall see, the OPT results for the quark number density are in very good agreement with the two fermion LQCD predictions. At the same time, the OPT results for the quark number susceptibilities agree well with the LQCD results up to about $1.2 \,T_c$ but behave just like the LN approximation (at $G_V \ne 0$) for higher temperatures. In particular, the coefficient $c_2$ which appears in the Taylor expansion for the pressure, $P/T^4=  c_0 + c_{2} (\mu/T)^{2}+ \dots$, presents a maximum at $T \sim 1.2 \, T_c$ that is not seem in any LQCD result. 
Very recently, Schramm and Steinheimer \cite {stefanGV2} and also Sugano et. al. \cite {sugano} have faced the same problem when employing the LN approximation to the PNJL at $G_V \ne 0$. The authors have then used this fact as a guide to understand how the vector interaction behaves.  Schramm and Steinheimer concluded  that there should be  a strong vector repulsion in the hadronic phase and near-zero repulsion in the deconfined phase while Sugano et al.  have estimated $G_V =  G_S \simeq 0.33 G_S$ (at $T=0$) by  requiring the entangled Polyakov--Nambu--Jona-Lasinio model (EPNJL) to fit  lattice QCD data obtained  with two-flavor Wilson fermions and large pion masses \cite{ejiri}.  A similar type of investigation, performed with a model which describes quarks and massive vector fields, has also beem performed by  Ferroni and Koch  \cite {lorenzo}.

 Understanding the origin of  $c_2(T)$  maximum presented by the OPT and the LNGv approximation is also one of our major goals. In this vein, we perform a simple high-$T$ exercise to identify its mathematical  origin before indicating the possible ways in which the discrepancy will be circumvented.   
This work is organized as follows.  We start by presenting  the PNJL model in the next section. Then, in Sec. III we implement the OPT and evaluate the free energy to the first non trivial order. Our numerical results are analyzed in Sec. IV and our final conclusions are presented in Sec. V. An appendix contains the details of the traces in color space.

\section{The Effective Quark Model}
In the case of effective quark theories in 3+1 dimensions the OPT was first applied \cite {prc,prc1} to the standard version of the NJL with the aim of studying how finite $N_c$ corrections influence the chiral transition pattern. However  the standard NJL model does not incorporate confinement and therefore is of limited interest if one aims to perform a  realistic description of QCD.
In this case, it becomes mandatory to find some way of simulating confinement within the original model.
With this purpose the \textit{Polyakov loop} has been added to the original NJL Lagrangian density \cite {njl} to produce the Polyakov-loop-extended 
NJL model (PNJL) \cite{pnjl}
\be
\mathcal{L}=\bar{\psi}\left(i\gamma_\mu D^\mu -\hat{m}_c\right)\psi+G_S\left[\left(\bar{\psi}\psi\right)^2
+\left(\bar{\psi}i\gamma_5\bolsy\tau\psi\right)^2\right]-G_V({\bar \psi} \gamma^\mu \psi)^2-\mathcal{U}\left(l,l^*,T\right), \label{PNJLlagrangian}
\ee
where $\psi$ (a sum over flavors and color degrees of freedom is
implicit) represents a flavor isodoublet ($u$ and $d$ type of quarks)
$N_{c}$-plet quark fields while $\vec \tau$ are isospin Pauli
matrices.  
The covariant derivative is given by 
\begin{align}
 D^\mu=\partial^\mu-iA^\mu \ \ \ \text{where} \ \ \ A^\mu=\delta^0_\mu A^0,
\end{align}
with the $SU(N)$ gauge coupling constant, $g$, absorbed in $A^\mu\left(x\right)=g\mathcal{A}^\mu_a\left(x\right){\lambda_a}/{2}$, while
$\mathcal{A}^\mu_a\left(x\right)$ represents the $SU\left(3\right)$ gauge field and $\lambda_a$ represent the Gell-Mann matrices.
Before presenting the Polyakov potential $\mathcal{U}\left(l,l^*,T\right)$ let us  define the Wilson line which winds once through a periodic time direction
\be
L\left(\bolsy x\right)\equiv \mathcal{P}\exp\left[i\int^\beta_0 d\tau A_4\left(\tau,\bolsy x\right)\right],
\ee
where $\beta=1/T$ and $A_4=iA_0$ is the temporal component of the Euclidean gauge field ($A_4,\bolsy A$).\par
The potential $\mathcal{U}\left(l,l^*,T\right)$ is fixed by comparison with pure-gauge lattice QCD \cite{claudia1}, 
from which one obtains the following ansatz \cite{claudia2},
\be
\dfrac{\mathcal{U}\left(l,l^*,T\right)}{T^4}=-\dfrac{1}{2}b_2\left(T\right)ll^*+b_4\left(T\right)\ln
\left[1-6ll^*+4\left(l^3+l^{*^3}\right)-3\left(ll^*\right)^2\right],\label{Upot}
\ee
with
\be
b_2\left(T\right)=a_0+a_1\left(\dfrac{T_0}{T}\right)+a_2\left(\dfrac{T_0}{T}\right)^2, \ \ b_4\left(T\right)=b_4\left(\dfrac
{T_0}{T}\right)^3\,.
\ee
The parameters values are $a_0=3.51$, $a_1=-2.47$, $a_2=15.22$, and $b_4=-1.75$ while
$T_0$ represents the critical temperature for deconfinement in the pure-gauge sector whose value is
fixed at 270 MeV \cite {claudia2}.\par
The expectation value of the Polyakov loop  is then given by \cite{fukushimapnjl}
\be
\Phi\equiv\left<l\left(\bolsy x\right)\right>, \ \ \ \text{and} \ \ \ \bar{\Phi}\equiv\left<l^*\left(\bolsy x\right)\right>,
\ee
where 
\begin{align}
l\left(\bolsy x\right)\equiv \dfrac{1}{N_c}\text{Tr}L\left(\bolsy x\right)
\end{align}

\section{Quark Pressure with Finite $N_c$ Contributions at vanishing $G_V$ }

Let us now use the OPT to evaluate the PNJL free energy beyond the large-$N_c$ limit.
The basic idea of this analytical nonperturbative method is to deform the original Lagrangian density by adding a quadratic term like  $(1-\delta)\eta \bar{\psi} \psi$ to the original Lagrangian density as well as by multiplying all coupling constants by $\delta$ \cite {prc}. The new parameter $\delta$ is just a bookkeeping label and $\eta$ represents an {\it arbitrary} mass parameter. Perturbative calculations are then performed in powers of the dummy parameter $\delta$ which is
formally treated as small and set to the original value, $\delta=1$,  at the end\footnote {Recall that within the large-$N_c$ on performs an expansion in powers of $1/N_c$ where $N_c$ is formally treated as large but set to the original value ($N_c=3$ in our case) at the end.}. Therefore, the fermionic propagator is dressed by  $\eta$ which may also be viewed as an infra red regulator in the case of massless theories. After a physical quantity such as the quark free energy density, $\cal F$, is evaluated to the order-$k$ and $\delta$ set to the unity a residual $\eta$ dependence  remains. Then,  optimal  nonperturbative results can be obtained by requiring that ${\cal F}_{\rm OPT}^{(k)}(\eta)$ be evaluated where it is less sensitive to variations of the arbitrary mass parameter. This requirement translates into the criterion known as the Principle of Minimal 
Sensitivity (PMS) \cite {pms}
\begin{equation}
\left. \frac{d{{\cal F}_{\rm OPT}}^{(k)}(\eta)}{d\eta }\right\vert _{\bar{\eta},\delta
  =1}=0\;.
\label{pms12}
\end{equation}
In general, the solution to this equation implies in self consistent relations  generating a nonperturbative coupling dependence.
In most cases nonperturbative   $1/N_c$ corrections appear  already at the first non trivial order while the large-$N_c$ (or MFA) results can be recovered at any time simply by considering $N_c \to \infty$. Finally, note that the OPT has the same
spirit as the Hartree and the Hartree-Fock approximation in which one also adds and subtracts a mass term. However, within these
two traditional approximations the topology of the dressing is fixed from the start: direct (tadpole) terms for Hartree and
direct plus exchange terms for Hartree-Fock. On the other hand, within the OPT, the dressed mass term ($\bar \eta$) acquires characteristics which change order by order progressively incorporating direct, exchange, vertex corrections, etc \footnote{These three different methods have been recently compared in Ref. \cite {ijmpe}.}.
To implement the OPT within the PNJL model at $G_V=0$ one can follow the prescription
used in Ref.~\cite{prc} basically replacing $m_c\rightarrow m_c+\left(1-\delta\right)\eta$ and $G_S\rightarrow \delta G_S$ in the PNJL Lagrangian density. According to this prescription
the deformed Lagrangian density for the PNJL model in terms of  auxiliary
fields, $\sigma$ and $\bolsy \pi$, becomes

\begin{equation}
\mathcal{L}_{\rm OPT}=\bar{\psi}\left[ i\gamma_\mu D^\mu-m_{c}-\eta \left(
  1-\delta \right)-\delta \left(
  \sigma +i{\gamma }_{5}{\bolsy \tau}\cdot {\bolsy \pi} \right)  \right] {\psi }-\delta \frac{1 }{4G_S }\left(
\sigma ^{2}+{\bolsy\pi }^{2}\right)-\mathcal{U}\left(l,l^*,T\right) .  
\label{delta12}
\end{equation}
\noindent
Note that since this model has a scalar as well as pseudo scalar channel the most general form for the mass parameter would have the form \cite{firstOPT}

\begin{equation}
\eta = \eta_1 +i{\gamma }_{5}{\bolsy \tau}\cdot {\bolsy \eta}_2,   
\label{eta12}
\end{equation}
\noindent
implying  four mass parameters, $\eta_1$ and the three
components of ${\bolsy \eta}_2$, to be fixed by the PMS.  However, for the  free energy only the fluctuations  in
the scalar direction become relevant once only the scalar field $\sigma$
acquires a non-zero  vacuum expectation value ($\langle \sigma \rangle\equiv {\bar \sigma} \ne 0$).  In other words, we can now assume  $ \langle {\bolsy \pi} \rangle =0$, which can be shown to imply that $\overline{{\bolsy \eta}_2}=0$ ~\cite{firstOPT}. Taking this
solution one needs to consider the simplest variational interpolation
involving only one  mass parameter, $\eta \equiv \eta_1$. Then, in order to evaluate $\cal F$ to the first non trivial  order one needs to consider the first two diagrams of Fig. 1. To better understand the method let us examine the different contributions in powers of $\delta$ as well as $N_c$ appearing to order-$\delta^2$ in Fig. 1. The first graph, which is of order-$\delta^0 N_c$, represents the usual one loop term  which also contributes to most approximations (e.g. Hartree, LN, etc). A diagram with this simple (direct) topology contributes to the free energy even when the system is composed by free fermions. The two loop contributions are of order $\delta N_c^0$ and represent exchange type of contributions which typically appear in a Hartree-Fock type of evaluation \cite {klevansky} or in a NLO $1/N_c$ evaluation. The third contribution, $O(\delta^2 N_c^0)$, brings a correction to the meson propagator and would also belong to the NLO in a usual $1/N_c$ evaluation. The last two contributions, which would belong to the NNLO in a $1/N_c$ expansion, bring the first vertex correction (forth graph) together another exchange correction to the quark propagator (fifth graph).  In summary the OPT perturbative character mixes up contributions which would belong to different orders within other approximations. A general outcome is that the LN result is readily recovered by taking the limit $N_c \to \infty$ in most OPT applications \cite {npb} as we shall explicitly show in the present PNJL case. At the same time the first non trivial $1/N_c$ corrections appear already at the first order and for the present case they display the topology of exchange like contributions. The explicit evaluation of the order-$\delta^2$ shown in Fig. 1 is beyond the scope of the present work but will be considered  in future applications. 

Then, considering the $\sigma$ direction only and applying   the Feynman rules to the two first diagrams of Fig. 1 one can write the order-$\delta$ OPT free energy as \cite {prc}

\begin{figure} 
\centering
 \includegraphics[width=.3\linewidth]{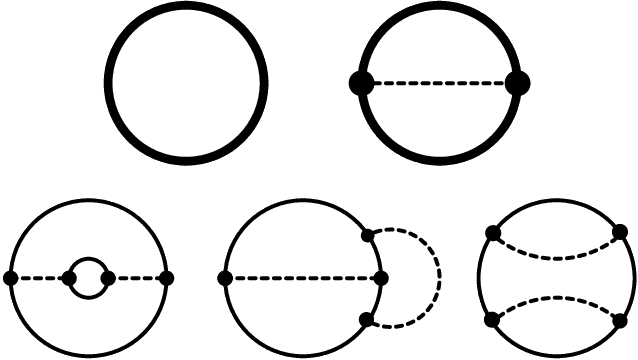} \quad \ 
 
 \caption{Radiative corrections to the free energy up to order-$\delta^2$. The first Feynman graph is of order-$\delta^0 \, N_c$, the second is of order-$\delta \, N_c^0$, the third is of order-$\delta^2 \, N_c^0$ while the fourth and the fifth are  of order-$\delta^2 \, N_c^{-1}$. The thick continuous lines represent the OPT fermionic dressed propagator, written in terms of $\eta^*=\eta+\delta(\sigma-\eta)$, which needs to be further expanded to $O(\delta^2)$.The thin continuous lines appearing in the last three contributions are written in terms of $\eta$ only since these graphs are already of second order in $\delta$. The dashed lines represent propagators associated to the background fields $\sigma$ and ${\bolsy \pi}$.}
  \label{diagrams}
\end{figure}

\begin{align}
  \mathcal{F}_{\rm OPT}&= \mathcal{U}\left(l,l^*,T\right)+\delta \frac{\sigma^2}{4G_S}+N_f 2i\int\frac{d^4 P}
 {\left(2\pi\right)^4}\mathrm{Tr_c}\ln\left[-P^2+\left(\eta^*+m_c\right)^2\right]\notag\\
 &-\delta G_S N_f 16\int\frac{d^4 P}{\left(2\pi\right)^4}\frac{d^4 Q}{\left(2\pi\right)^4}\mathrm{Tr_c}
 \frac{P_0}{\left [-P^2+\left(\eta+m_c\right)^2\right ]}\frac{Q_0}{\left [-Q^2+\left(\eta+m_c\right)^2\right ]}\notag\\
 &+\delta G_S N_f\left(\eta+m_c\right)^2 8 \int\frac{d^4 P}{\left(2\pi\right)^4}\frac{d^4 Q}{\left(2\pi\right)^4}\mathrm{Tr_c}
 \frac{1}{\left [-P^2+\left(\eta+m_c\right)^2\right ]}\frac{1}{\left [-Q^2+\left(\eta+m_c\right)^2\right ]} + O(\delta^2)\,,\label{319}
\end{align}
where $\eta^*= \eta+\delta (\sigma-\eta)$, $P=(p_0-iA_4,{\bolsy p})$,  and $\mathrm{Tr_c}$ indicates the trace over color space.
In order to introduce the control parameters $T=1/\beta$ and $\mu$ one can use the Matsubara's (imaginary time) formalism  
\begin{align}
 \int\frac{d^4 p} {\left(2\pi\right)^4}\rightarrow \frac{i}{\beta} \sum_n \int \frac{d^3p}{\left(2\pi\right)^3}\label{intomega}  \,\,,
\end{align}
where the quadrimomentum is given as $p=\left(i\omega_{n}+\mu-iA_4,{\bolsy p}\right)$, while  the Matsubara frequencies for fermions are given by $\omega_n=\left(2n+1\right)/\beta$, $n=0,\pm 1,\pm 2\ldots$.  Denoting $\mu^\prime=\mu-iA_4$ and expanding $\eta^*$ to order-$\delta$ one gets
\begin{align}
  \mathcal{F}_{\rm OPT}&=\mathcal{U}\left(l,l^*,T\right)+\delta \frac{\sigma^2}{4 G_S}-\frac{N_f}{\beta}2 \int\frac{d^3 p}
 {\left(2\pi\right)^3}\mathrm{Tr_c}\sum_n\ln\left[\left(\omega_{n}-i\mu^\prime\right)^2+E_p^2\right]\notag\\
 &-\delta\frac{N_f}{\beta}\left(\eta+m_c\right)\left(\eta-\sigma\right)4 \int\frac{d^3 p}{\left(2\pi\right)^3}\mathrm{Tr_c}\sum_n\frac{1}
 {\left [\left(\omega_n-i\mu^\prime\right)^2+E_p^2\right ]}\notag\\
 &-\delta\frac{G_S N_f}{\beta^2}16 \int\frac{d^3 p}{\left(2\pi\right)^3}\frac{d^3 p}{\left(2\pi\right)^3}\mathrm{Tr_c}\sum_n
 \sum_m\frac{\omega_n-i\mu^\prime}{\left [\left(\omega_n-i\mu^\prime\right)^2+E_p^2\right ]}\frac{\omega_m-i\mu^\prime}{\left [ \left(\omega_m-i\mu^\prime\right)^2+E_q^2\right ]}\notag\\
 &+\delta\frac{G_S N_f}{\beta^2}\left(\eta+m_c\right)^2 8 \int\frac{d^3 p}{\left(2\pi\right)^3}\frac{d^3 p}{\left(2\pi\right)^3}
\mathrm{Tr_c}\sum_n\sum_m
 \frac{1}{\left [\left(\omega_n+i\mu^\prime\right)^2+E_p^2\right ]}
 \frac{1}{\left [ \left(\omega_m+i\mu^\prime\right)^2+E_q^2\right ]} +O(\delta^2),\label{320}
\end{align}
where the dispersion is $E_p^2=[{\bolsy p}^2+(m_c+\eta)^2]$. Note that the expansion of $\eta^*$ contained in the one loop contribution in Eq. (\ref{319}) automatically generates the two order-$\delta$ (one loop) contributions which are contained in the $(\eta-\sigma)I_2$ term of Eq. (\ref{320}).
 To our knowledge the color trace evaluation has 
not been performed before for the two loop contributions represented by the two last terms of Eq. (\ref{320}) and therefore we present this  straightforward,
but lengthly, exercise in Appendix \ref{AppendixA}. Then, one finally obtains
\begin{align}
 \mathcal{F}_{\rm OPT}\left(\eta,\sigma,l,l^*,\mu,T\right)= & \mathcal{U}\left(l,l^*,T\right)+\frac{\sigma^2}{4G_S}-2N_fI_1\left(\mu,T\right)+\delta 2 N_fN_c\left(\eta+m_c\right)\left(\eta-\sigma\right)I_2\left(
 \mu,T\right)\notag\\
 &+ \delta 4 G_SN_fN_c\left[I^2_3\left(\mu,T\right)+\Delta I_3\left(\mu,T\right)\right]\notag\\
 &- \delta 2 G_SN_fN_c\left(\eta+m_c\right)^2\left[I^2_2\left(\mu,T\right)+\Delta I_2\left(\mu,T\right)\right] +O(\delta^2),\label{freenergyOPT}
\end{align}
where $I_i(\mu,T)$ $(i=1,2,3)$ represent the following integrals
\begin{align}
  I_1\left(\mu,T\right)=&\int\frac{d^3 p}{\left(2\pi\right)^3×}\left\lbrace N_cE_p+\ln\left[g^+_l\left(E_p\right)\right]
  +\ln\left[g^-_l\left(E_p\right)\right]\right\rbrace,\label{322}\\
  I_2\left(\mu,T\right)=&\int\frac{d^3 p}{\left(2\pi\right)^3×}\frac{1}{E_p×}\left[1-f^+_l-f^-_l\right],\\
  I_3\left(\mu,T\right)=&\int\frac{d^3 p}{\left(2\pi\right)^3×}\left[f^+_l-f^-_l\right]\label{I3} \,.
\end{align}
Here, we impose the cutoff only for the vacuum terms (the two first contributions on the right hand side of $I_1$ and $I_2$), since the thermal contribution  has a natural cutoff in itself specified
by the temperature \cite {reg,prc}. In the presence of the Polyakov loop the fermionic distribution functions read
\begin{align}
 f^+_{l}\left(E_p\right)&=\frac{l e^{-\beta\left(E_p-\mu\right)}+2l^* e^{-2\beta\left(E_p-\mu\right)}+e^{-3\beta\left(E_p-\mu\right)}}
 { g^+_l\left(E_p\right)×},\\
  f^-_{l}\left(E_p\right)&=\frac{l^* e^{-\beta\left(E_p+\mu\right)}+2l e^{-2\beta\left(E_p+\mu\right)} +e^{-3\beta\left(E_p+\mu\right)}}
  {g^-_l\left(E_p\right)×},\\
  g^+_l\left(E_p\right)&=1+3l e^{-\beta\left(E_p-\mu\right)}+3l^* e^{-2\beta\left(E_p-\mu\right)}+  e^{-3\beta\left(E_p-\mu\right)},\\
  g^-_l\left(E_p\right)&=1+3l^* e^{-\beta\left(E_p+\mu\right)}+3l e^{-2\beta\left(E_p+\mu\right)} +e^{-3\beta\left(E_p+\mu\right)}.
\end{align}
Finally, the thermal integrals $\Delta_2$ and $\Delta_3$ which contribute to the two loop contributions only (see Appendix) are given by
\begin{align}
\Delta I_3\left(\mu,T\right)=\int\frac{d^3 p}{\left(2\pi\right)^3}\frac{d^3 q}{\left(2\pi\right)^3}\Delta_3,
\end{align}
and
\begin{align}
\Delta I_2\left(\mu,T\right)=\int\frac{d^3 p}{\left(2\pi\right)^3}\frac{d^3 q}{\left(2\pi\right)^3}\Delta_2,
\end{align}
where
\begin{align}
 \Delta_2&=\frac{e^{-\beta\left(E_p-\mu\right)}e^{-\beta\left(E_q-\mu\right)}}{g^+_l(E_p)g^+_l(E_q)}\left\lbrace2\left(l^2-l^*\right)+\left(ll^*-1\right)
 \left[e^{-\beta\left(E_p-\mu\right)}+e^{-\beta\left(E_q-\mu\right)}\right]\right.\notag\\
 &+\left.2\left({l^*}^2-l\right)e^{-\beta\left(E_p-\mu\right)}e^{-\beta\left(E_q-\mu\right)}\right\rbrace\notag\\
 &+\frac{e^{-\beta\left(E_p-\mu\right)}e^{-\beta\left(E_q+\mu\right)}}{g^+_l(E_p)g^-_l(E_q)}\left\lbrace2\left(l-{l^*}^2\right)
 e^{-\beta\left(E_p-\mu\right)}+2\left(l^*-l^2\right) e^{-\beta\left(E_q+\mu\right)}\right.\notag\\
 &+\left.\left(1-ll^*\right)\left[1+e^{-\beta\left(E_p-\mu\right)}e^{-\beta\left(E_q+\mu\right)}
 \right]\right\rbrace\notag\\
 &+\frac{e^{-\beta\left(E_q-\mu\right)}e^{-\beta\left(E_p+\mu\right)}}{g^+_l(E_q)g^-_l(E_p)}\left\lbrace2\left(l-{l^*}^2\right)
 e^{-\beta\left(E_q-\mu\right)}+2\left(l^*-l^2\right) e^{-\beta\left(E_p+\mu\right)}\right.\notag\\
 &+\left.\left(1-ll^*\right)\left[1+e^{-\beta\left(E_q-\mu\right)}e^{-\beta\left(E_p+\mu\right)}
 \right]\right\rbrace\notag\\
 &+\frac{e^{-\beta\left(E_p+\mu\right)}e^{-\beta\left(E_q+\mu\right)}}{g^-_l(E_p)g^-_l(E_q)}\left\lbrace2\left({l^*}^2-l\right)+\left(ll^*-1\right)
 \left[e^{-\beta\left(E_p+\mu\right)}+e^{-\beta\left(E_q+\mu\right)}\right]\right.\notag\\
 &+\left.2\left(l^2-l^*\right)e^{-\beta\left(E_p+\mu\right)}e^{-\beta\left(E_q+\mu\right)}\right\rbrace,
\end{align}
and
\begin{align}
 \Delta_3&=\frac{e^{-\beta\left(E_p-\mu\right)}e^{-\beta\left(E_q-\mu\right)}}{g^+_l(E_p)g^+_l(E_q)}\left\lbrace2\left(l^2-l^*\right)+\left(ll^*-1\right)
 \left[e^{-\beta\left(E_p-\mu\right)}+e^{-\beta\left(E_q-\mu\right)}\right]\right.\notag\\
 &+\left.2\left({l^*}^2-l\right)e^{-\beta\left(E_p-\mu\right)}e^{-\beta\left(E_q-\mu\right)}\right\rbrace\notag\\
 &-\frac{e^{-\beta\left(E_p-\mu\right)}e^{-\beta\left(E_q+\mu\right)}}{g^+_l(E_p)g^-_l(E_q)}\left\lbrace2\left(l-{l^*}^2\right)
 e^{-\beta\left(E_p-\mu\right)}+2\left(l^*-l^2\right) e^{-\beta\left(E_q+\mu\right)}\right.\notag\\
 &+\left.\left(1-ll^*\right)\left[1+e^{-\beta\left(E_p-\mu\right)}e^{-\beta\left(E_q+\mu\right)}
 \right]\right\rbrace\notag\\
 &-\frac{e^{-\beta\left(E_q-\mu\right)}e^{-\beta\left(E_p+\mu\right)}}{g^+_l(E_q)g^-_l(E_p)}\left\lbrace2\left(l-{l^*}^2\right)
 e^{-\beta\left(E_q-\mu\right)}+2\left(l^*-l^2\right) e^{-\beta\left(E_p+\mu\right)}\right.\notag\\
 &+\left.\left(1-ll^*\right)\left[1+e^{-\beta\left(E_q-\mu\right)}e^{-\beta\left(E_p+\mu\right)}
 \right]\right\rbrace\notag\\
 &+\frac{e^{-\beta\left(E_p+\mu\right)}e^{-\beta\left(E_q+\mu\right)}}{g^-_l(E_p)g^-_l(E_q)}\left\lbrace2\left({l^*}^2-l\right)+\left(ll^*-1\right)
 \left[e^{-\beta\left(E_p+\mu\right)}+e^{-\beta\left(E_q+\mu\right)}\right]\right.\notag\\
 &+\left.2\left(l^2-l^*\right)e^{-\beta\left(E_p+\mu\right)}e^{-\beta\left(E_q+\mu\right)}\right\rbrace.
\end{align}

Having presented the mathematical expressions  let us now discuss the physics related to the
 OPT free energy, Eq. (\ref{freenergyOPT}) so that one may gain an intuitive insight about the expected results. 
 The first term  contained in ${\cal F}_{\rm OPT}$ represents 
 the classical potential while the second is similar to the standard result obtained in
the case of free fermionic gas whose masses are given by $m_c+\eta$ as $I_1$ suggests. The terms proportional to $I_2 \sim
\partial I_1/\partial\eta$ are reminiscent of the one loop scalar density, $\rho_s = \left\langle\bar\psi\psi\right\rangle$. At the same time, the terms proportional to
$I_3 \sim \partial I_1/\partial\mu$ only survive when $\mu \ne 0$ as Eq. (\ref{I3}) shows. This can be easily understood by recalling that, to one loop,
the quark number density $\rho_q=\left\langle\psi^+\psi\right\rangle$ is given by $I_3$. Then, by noting that $I_3$ is $1/N_c$ suppressed one can readily draw
the basic physical differences between the OPT and LN approximation at this first non-trivial order. Namely, the OPT free energy is written in terms of scalar and vector condensates while only the scalar density contributes to the latter. Therefore, at least to the first non trivial order one can expect that the finite $N_c$ corrections  will be more pronounced at finite densities as Refs. \cite {prdgn3d,prc,prc1} suggest. This
 is an important observation for the discussions to be carried out in the sequel.\par
In order to obtain thermodynamical quantities from the OPT free energy one must  consider the following set of coupled equations 
\begin{align}
 \left.\frac{\partial\mathcal{F}_{\rm OPT}}{\partial \eta×}\right|_{\bar\eta}=0, \ \ \  \left.\frac{\partial\mathcal{F}_{\rm OPT}}{\partial \sigma×}\right|_{\bar\sigma}=0, \ \ \  \left.\frac{\partial\mathcal{F}_{\rm OPT}}{\partial \l×}\right|_{\Phi}=0,\ \ \  
 \left.\frac{\partial\mathcal{F}_{\rm OPT}}{\partial \l^*×}\right|_{\bar\Phi}=0, \label{333}
\end{align}
which can be solved numerically.
In order to illustrate how the OPT generates nonperturbative results from a purely perturbative evaluation it is convenient to consider  the analytical form of the PMS equation (first one in Eq. \ref{333}) 
 \begin{align}
 \left\lbrace\left[\eta-\sigma-2\left(\eta+m_c\right)G_SI_2\right]\left[1+\left(\eta+m_c\right)\frac{\partial}{\partial \eta}\right]I_2+4G_SI_3
 \frac{\partial I_3}{\partial \eta×}+2G_S\frac{\partial \Delta I_3}{\partial \eta×}-G_S\left(\eta+m_c\right)^2\frac{\partial \Delta I_2}{\partial \eta×}\right\rbrace_{\eta=\bar\eta}=0.\label{PMSeta}
\end{align}
To make the optimization process even more transparent let us consider, just for the moment, the large-$N_c$ limit. In this situation all terms proportional to $G_S$ would be neglected yielding 
\begin{align}
 \left\lbrace\left[\eta-\sigma\right]\left[I_2+\left(\eta+m_c\right)\frac{\partial I_2 }{\partial \eta}\right]\right \rbrace_{\eta=\bar\eta}=0.\label{PMSetaLN}
\end{align}
The first term gives the simple solution $\bar \eta = \sigma$ which exactly reproduces the large-$N_c$ result as one can easily check by discarding the $1/N_c$ suppressed contributions represented by the two last terms appearing in Eq. (\ref {freenergyOPT}). As discussed in Ref. \cite {npb} the second solution (which depends only upon energy scales) is unphysical. Plugging the solution $\bar \eta = \sigma$ into the gap equation 
 (second one in Eq. \ref{333}) 
\begin{align}
 \bar\sigma=4G_S N_fN_c\left(\eta+m_c\right)I_2\label{sigma},
\end{align}
one exactly retrieves  the familiar LN results. Of course here we are considering the realistic $N_c=3$ case so that the PMS equation is not so simple but nevertheless it is reassuring that the OPT easily reproduces the LN  ``exact"  result when the limit $N_c \to \infty$ is taken.  
Finally, to obtain the  OPT pressure for the PNJL case one simply considers
\begin{align}
  P_{\rm OPT}=-\mathcal{F}_{\rm OPT}\left(\bar\eta,\bar\sigma,\Phi,\bar\Phi,\mu,T\right).
\end{align}

\section{Large-$N_c$ Quark Pressure at  with Finite $G_V$ Contributions }

In the previous section we have explicitly show how finite $N_c$ radiative corrections can generate the appearance of   density dependent terms which are absent in large-$N_c$ evaluations when $G_V=0$. Nevertheless, these important contributions can also be considered within the  LN framework if one modifies the original PNJL Lagrangian density by adding a repulsive vector term with strength $G_V$ as Eq. (\ref {PNJLlagrangian}) shows.  In this case  the large-$N_c$ (LNGv)  free energy evaluation  within this model is standard and yields \cite {stefanGV} 
\begin{align}
{\cal F}_{\rm LNGv}\left(\sigma,\Phi,{\bar \Phi},\mu,T\right)=\mathcal{U}\left(l,l^*,T\right)+\frac{(M_{\rm LN}-m_c)^2 }{4G_S}
 -4I_1\left(M_{\rm LNGv},{\tilde\mu}_{\rm LNGv},T\right)-4G_VN_f^2N_c^2I_3^2\left(M_{\rm LNGv},{\tilde \mu}_{\rm LNGv},T\right),\label{272}
\end{align}
 where
\begin{equation}
M_{\rm LNGv}= m_c - 2G_S \rho_s \,\,,
\end{equation}
and
\begin{equation}
{\tilde \mu}_{\rm LNGv} = \mu - 2G_V \rho_q \,\, .
\end{equation}
Note that the above equations have been written in terms of  the quark number density and the scalar density which are
respectively given by
\begin{equation}
\rho_q = \langle  {\psi}^+ \psi \rangle = 2N_f N_c
 I_3(M_{\rm LNGv},T,{\tilde \mu_{\rm LNGv}})\,\,\,,
\label{rhoqLN}
\end{equation}
and
\begin{equation}
\rho_s = \langle  {\bar \psi} \psi \rangle = - 2N_f N_c  I_2(M_{\rm LNGv},T,{\tilde \mu_{\rm LNGv}})\,\,\,.
\label{rhosLN}
\end{equation}
Then, the self consistent equations for $M_{\rm LNGv}$ and ${\tilde \mu_{\rm LNGv}}$ have to solved 
together with
\begin{align}
  \left.\frac{\partial\mathcal{F}_{\rm LN}}{\partial \l}\right|_{\Phi}=0,\ \ \  
 \left.\frac{\partial\mathcal{F}_{\rm LN}}{\partial \l^*}\right|_{\bar\Phi}=0, \label{333}
\end{align}
to yield the  LN pressure, $P_{LN}=-{\cal F}_{LN}$.
As discussed in Ref. \cite {Fukushima}, together with ${\tilde \mu}_{LN}$,  the term $-4G_VN_f^2N_c^2I_3^2$ appearing in the LNGv free energy above produces a {\it net} effect proportional to $4G_VN_f^2N_c^2I_3^2$ which, upon replacing $G_V \to G_S/(N_c N_f)$, reproduces the OPT term $4G_S N_f N_c I_3^2$. Therefore, from a qualitative point of view   the same type of physics may be expected to arise within the two different approximations considered so far. Obviously, in the absence of a vector channel the LN results can be directly obtained from the above equations simply by setting $G_V=0$.

\section{Numerical results}
Let us now compare the numerical results obtained by using the different analytical approximations with those furnished by LQCD. Following Ref. \cite {sugano} we shall mainly consider the LQCD results obtained by Ejiri et al. \cite {ejiri} with two-flavor Wilson fermions and large pion masses but for completeness, in the evaluation of $c_2$, we will also consider the continuum extrapolated lattice QCD results obtained by Borsanyi et al. \cite {latticecont} at physical pion masses. With this aim we have defined two adequate parametrizations for  each approximation as table I displays. As the table also shows the chiral transition temperature, $T_\sigma$, and the confinement transition temperature, $T_\Phi$, are approximately the same in both cases so that for simplicity we set $T_c \equiv T_\sigma \simeq T_\Phi$ in our plots. Note also that here we do not impose $\Phi = {\bar \Phi}$ since this equality only holds when $\mu=0$.  At finite densities the LQCD simulations can make predictions if one considers the following Taylor expansion for the pressure
\begin{table}[h!]
\caption{Parameter sets for the OPT and for the LN approximation. Apart from $G_V=G_S/N_c$ the parametrizations for the LNGv and the LN approximations are exactly the same. } 
\begin{center}
\begin{tabular}{l || c | c | c |c | c | c} 
       \hline
       \hline
	& $m_\pi$[MeV] &  $\Lambda$ [MeV] & $m_c$ [MeV] & $G_S\Lambda^2$ & $T_\sigma$ [MeV] & $T_\Phi$ [MeV] \\
	\hline
	OPT& 500  & 590 & 72.3 & 1.91 & 221 & 220  \\
	  & 135 & 640 & 4.9 & 1.99 & 217 &213\\
	  \hline
	LN& 500 & 631.5 & 72 & 2.19 & 225 & 224 \\ 
	& 135& 631.5 & 5.5 & 2.19 & 225 & 215\\
	\hline
	\hline
\end{tabular}\label{tab:tabla4}
\end{center}
\end{table}
\begin{align}
 \frac{P}{T^4}=\sum_{n=0}^\infty c_{n}\left(T\right)\left(\mu/T\right)^{n} \,\,,
 \label{press}
\end{align}
where, due to the reflexion symmetry $P\left(\mu\right)=P\left(-\mu\right)$, only even powers of $\mu/T$ contribute. The coefficients of this series are of particular interest in the study of phase transitions since they are related to the   quark number susceptibilities. 
Once the pressure has been evaluated within a given model approximation they can be obtained from
\begin{align}
 c_n\left(T\right)=\frac{1}{n!}\left.\frac{\partial^n P\left(T,\mu\right)/T^4}{\partial \left(\mu/T\right)^n}\right|_{\mu=0}.\label{coeficientsNJL}
\end{align}
Having determined some of the $c_n$ in a LQCD evaluation one can further obtain other thermodynamical quantities like the quark number density or the quark number susceptibility
which are respectively given by
\begin{equation}
 \frac{\rho_q}{T^3}=\frac{\partial P\left(T,\mu\right)/T^4}{\partial \left(\mu/T\right)×}=2c_2\frac{\mu}{T×}+4c_4\left(\frac{\mu}{T}\right)^3
 +\cdots\\
 \end{equation}
 and
 \begin{equation}
 \frac{\chi_q}{T^2}=\frac{\partial^2 P\left(T,\mu\right)/T^4}{\partial \left(\mu/T\right)^2×}=2c_2+12c_4\left(\frac{\mu}{T}\right)^2
 +\cdots
\end{equation}
Let us start by comparing the quark number density  density, $\rho_q$, as a  function of $T$ for different values of $\mu$.  Figure 2 shows this quantity obtained from the {\it full} analytical expressions for the pressure as given by the OPT, the LN at $G_V=0$ and the LNGv at $G_V=G_S/N_c$  which is the value recently proposed in Ref. \cite {sugano}. All the  results provided by the different analytical approximations  are compared with the LQCD simulations for the two flavor case \cite {ejiri}. As one can see the OPT and the LNGv display similar results and are in good agreement with the LQCD predictions (especially at higher $\mu$ values) showing the importance of the repulsive vector channel in this case. The results for the quark number susceptibility, $\chi_q$,  are illustrated in Fig. 3. Again, in this case the standard LN approximation seems to miss important information as $\mu$ increases as the figure shows. On the other hand the OPT is in good agreement with LQCD simulations up to $T \sim 1.2 \, T_c$ but then, at higher temperatures, the drop of $\chi_q$   with $T$ is more pronounced within the former method.   It is interesting to remark, in the same figure, the opposite high-$T$ behavior displayed by the OPT  and the LN approximation  which hints to the fact that, at high-$T$, the dominating  OPT corrections to $\chi_q$ seem have a negative sign. Let us now recall that   at vanishing densities the quark number susceptibility is related to the coefficient $c_2$ appearing in the Taylor expansion of the pressure, Eq. (\ref {press}), and therefore it may also be instructive to compare  the different predictions for this quantity. This is done in Fig. 4 where one can see that the OPT results are in better agreement with the LQCD predictions up to $T \approx 1.2\, T_c$ than the LN (at $G_V=0$ and $G_V=G_S/N_c$). However, at $T \approx 1.3 \,T_c$ the OPT and the LNGv display a maximum which is not seem by the LQCD nor by the LN approximation at $G_V=0$.  For higher $T$ both the OPT and the LNGv continue to deviate from the Stefan--Boltzmann limit as well as from the LQCD data obtained in two different simulations \cite{ejiri,latticecont}. This discrepant behavior has been originally remarked by Schramm and Steinheimer who applied the large $N_c$ approximation  to the two flavor PNJL,  with $G_V \ne 0$, to evaluate the  second and fourth order quark number susceptibilities at zero baryochemical potential \cite {stefanGV}. The authors have interpreted this discrepancy as an indication   that that above $T_c$ any mean field type of repulsive vector interaction can be excluded from model calculations. Recently, the same authors have extended their analysis to the three flavor case basically reaching the same conclusions \cite {stefanGV2}. 
\begin{figure} 
\begin{subfigure}

\centering
 \includegraphics[width=.4\linewidth]{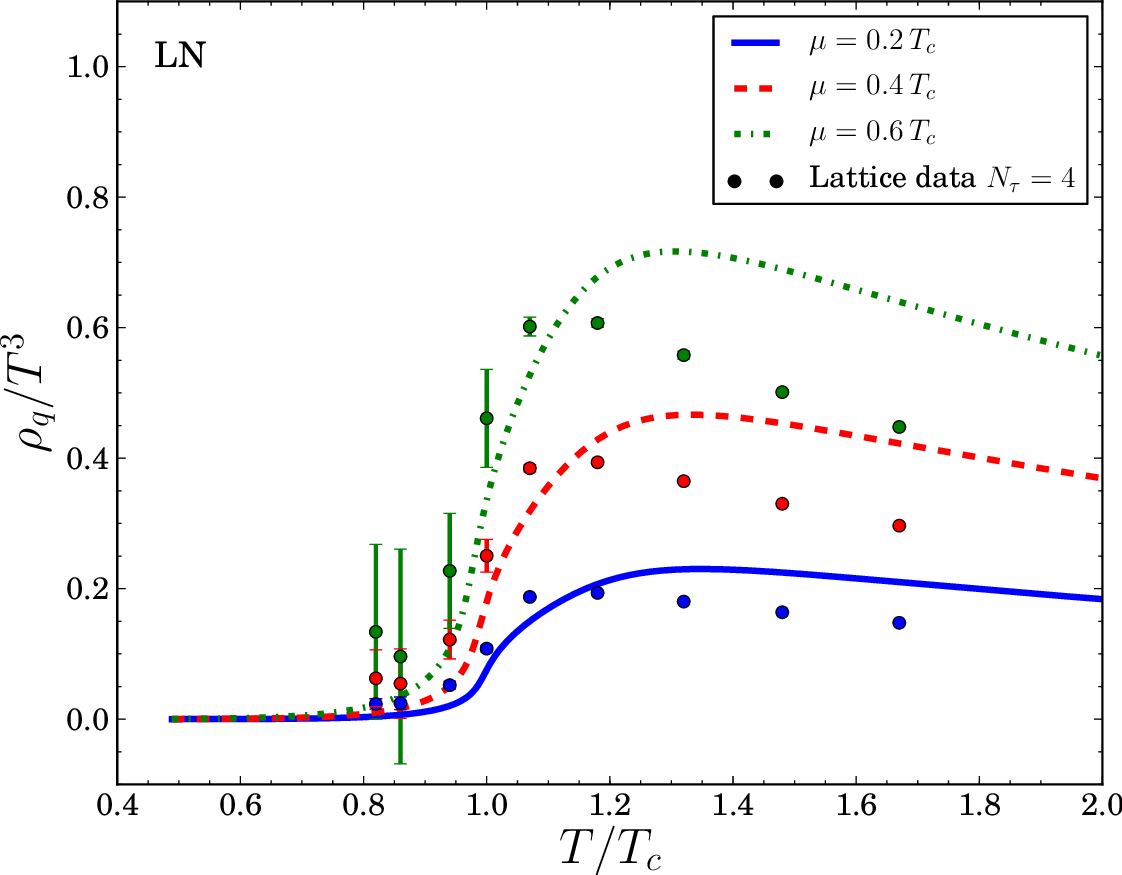} \quad \ 
 \end{subfigure}
 \begin{subfigure}
 \centering
 \includegraphics[width=.4\linewidth]{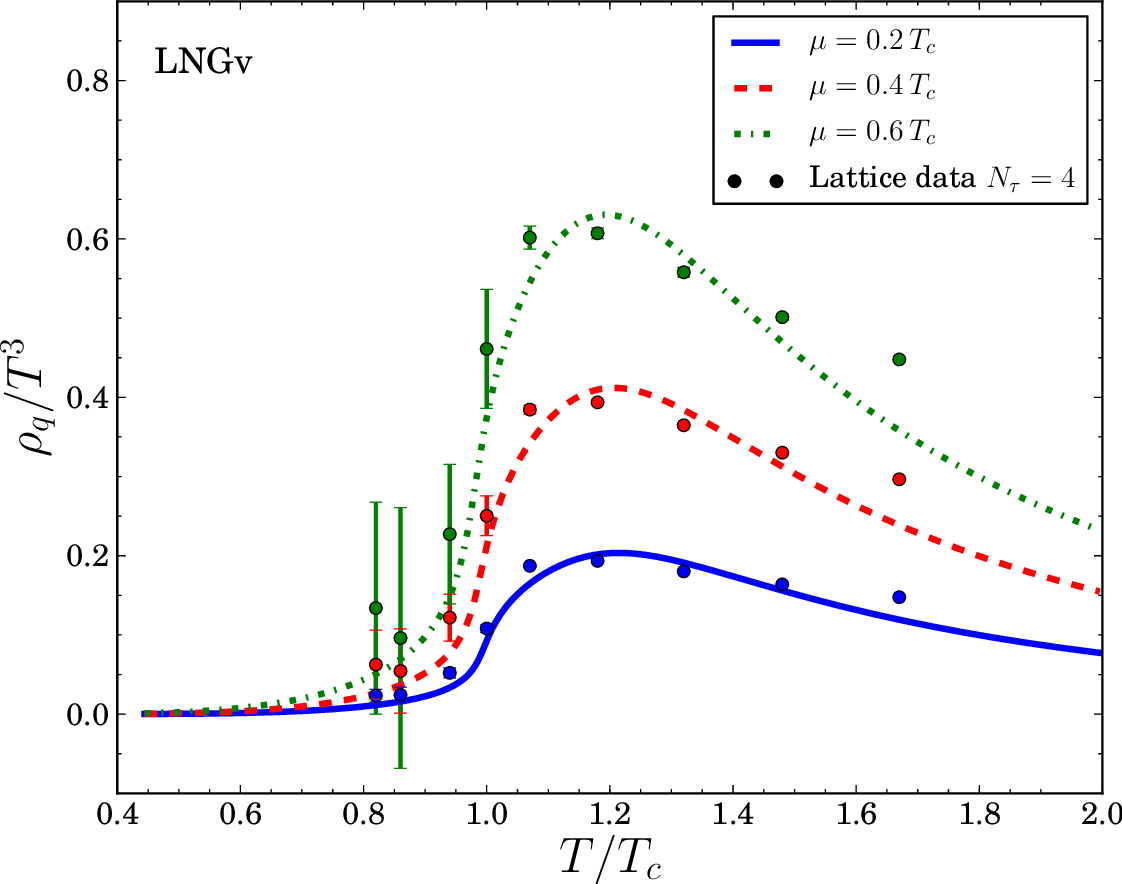} \quad \ 
 \end{subfigure}
 \begin{subfigure}
 \centering
 \includegraphics[width=.4\linewidth]{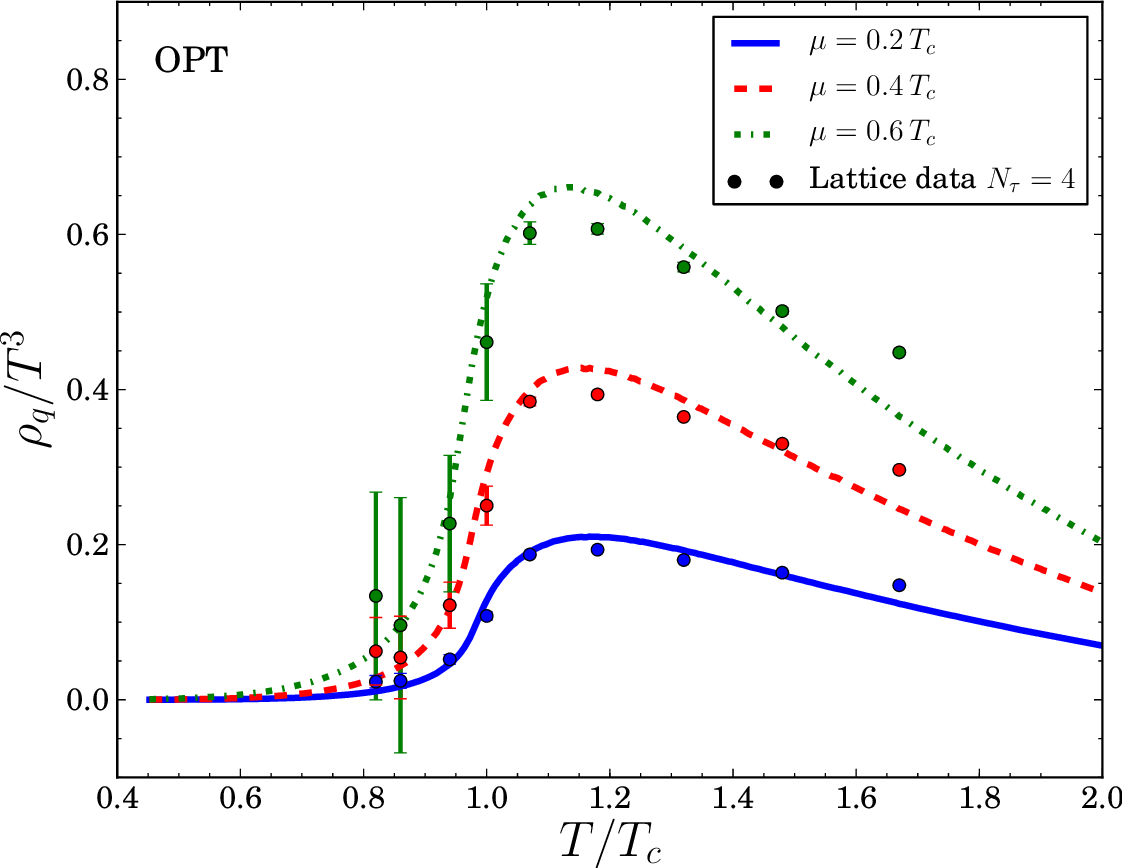} \quad \ 
 \end{subfigure}
 \caption{Normalized quark number density as a function of $T/T_c$ for different values of $\mu$ as predicted by the LN approximation (top left panel), the 
 LNGv approximation  (top right panel) and the OPT (bottom panel). The model parameters are for $m_\pi=500\,{\rm MeV}$ and the lattice data were taken from Ref. \cite{ejiri}.}
  \label{c2PNJL}
\end{figure}
However, we are explicitly showing that even when $G_V=0$ the coefficient $c_2$ has a peculiar behavior beyond the large-$N_c$ limit and,  since we are already working at $G_V=0$, we cannot conciliate the  OPT  results with those furnished by LQCD  by requiring  $G_V \to 0$ at $ T > T_c$. In order to understand the mathematical origin of the maximum displayed by $c_2$ let us examine how this quantity behaves, in the OPT case, at high temperatures. Using $\partial^2 P/\partial \mu^2$ at $\mu=0$ and high-$T$ (${\bar \Phi},\Phi \to 1$ and $\bar \sigma, \bar \eta \to 0$) one can easily verify that (see also Fig. 5)
\begin{equation}
  c_2\sim c_2^{LN} -8\delta G_SN_fN_c\left(\frac{\partial I_3}{\partial \mu}\right)^2 (2T^2)^{-1}+O(\delta^2)  \;\;,
  \label{c2}
  \end{equation}
 where
 \begin{equation}
  c_2^{LN}=2N_fN_c\left (\frac{\partial I_3}{\partial \mu}\right ) (2T^2)^{-1}\;\;.
 \end{equation}
The above expressions show that the dynamically generated repulsive vector  term gives a negative contribution to $c_2$ producing the observed maximum. Note that  the same observation applies to  LNGv approximation (in this case $G_S$ is replaced by $G_V N_c N_f$). On the other hand, in a  large-$N_c$ evaluation, at $G_V=0$, the scalar coupling  is of order $1/N_c$ and therefore the negative contribution is suppressed. At this point we can summarize our results as follows. First, the results for $\rho_q$ and $\chi_q$ show that the traditional LN approximation with $G_V=0$ will not provide accurate results at increasing densities but at the same time, by examining its predictions for $c_2$, it looks like this approximation performs well   at high temperatures where it quickly converges to the SB limit. As we have discussed, the LN problem in dealing with the high density domain can be solved either by explicitly introducing a repulsive vector channel at the classical level (still within the large-$N_c$ limit) or by evaluating radiative $1/N_c$ corrections. Then, for $T \lesssim 1.2 \, T_c$, the LQCD results for $\mu \ne 0$ will be more faithfully described but convergence towards the SB limit  will be lost. A glance at Eq. (\ref{c2}) suggests that another possibility is that higher order contributions may produce a contribution with a positive sign so that eventually $c_2$ will converge to the SB limit.  Within this scenario it appears that the (rather quick) convergence of the LN result for $c_2$ towards the SB could be accidental \footnote{Fluctuation corrections to the mean field approximation do not seem to generate any kind of vector repulsion and  the rise in $c_2$ is even sharper than the one observed  with the LN approximation \cite {ratti}.}.   As already emphasized, at order-$\delta^2$ the OPT considers three different topologies, including vertex corrections, which would belong to the NLO and NNLO in a $1/N_c$ expansion. Although the explicit evaluation of the cumbersome $\delta^2$ contributions is beyond the scope of the present work  we may, nevertheless, expect that higher order terms will end up   dressing the scalar coupling so that it $G_S$ will decrease with the temperature after $T_c$, just as the QCD coupling, so that $c_2$ end up by displaying the expected LQCD behavior. In the context of the LNGv approximation  Sugano et al. have invoked the possible thermal dependence of   $G_V$  as a way to explain the observed $c_2$ discrepancy but our investigation shows that the problem will arise even in the absence of $G_V$ when the first finite $N_c$ corrections are included. Therefore, also in this case it appears that thermal effects on the scalar coupling $G_S$ are important to reproduce the free fermion gas high-$T$ behavior. To verify this situation  let us now follow Ref. \cite {sugano} by considering the entangled PNJL model (EPNJL).

\begin{figure} 
\begin{subfigure}

\centering
 \includegraphics[width=.4\linewidth]{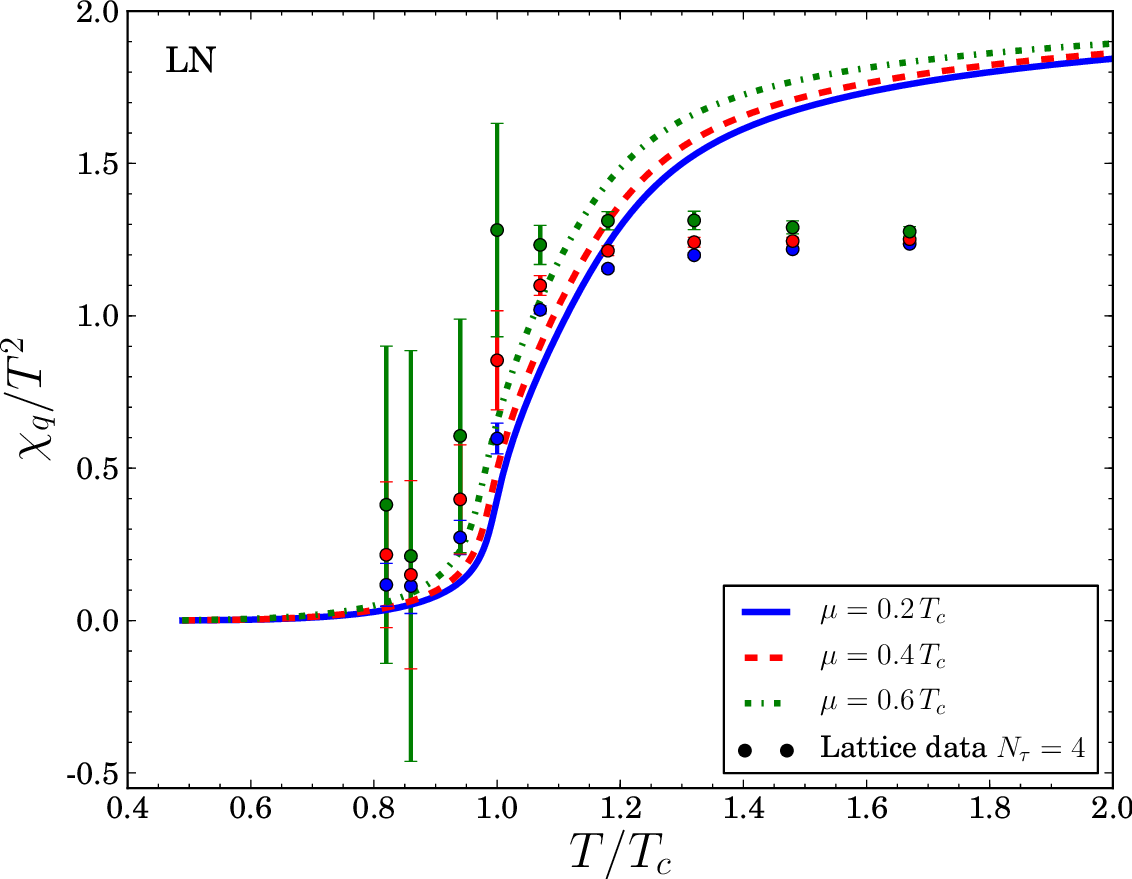} \quad \ 
 \end{subfigure}
 \begin{subfigure}
 \centering
 \includegraphics[width=.4\linewidth]{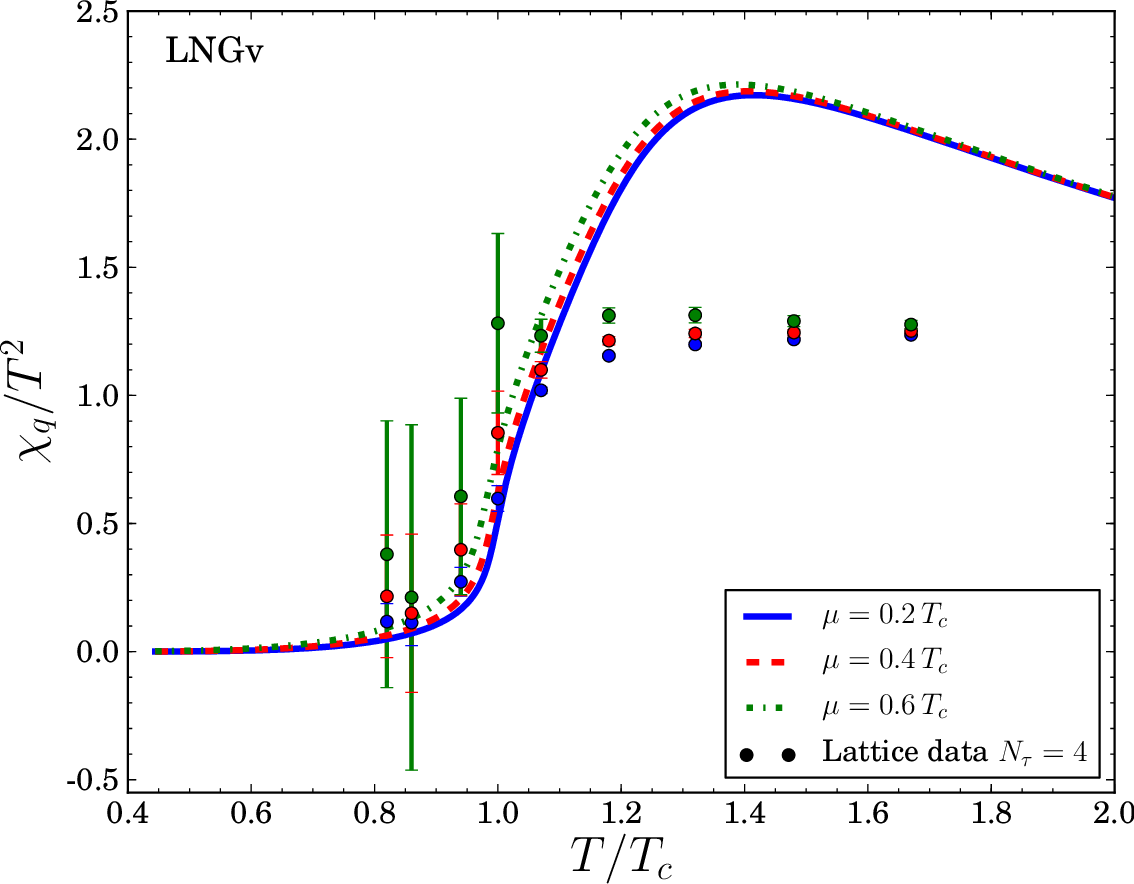} \quad \ 
 \end{subfigure}
 \begin{subfigure}
 \centering
 \includegraphics[width=.4\linewidth]{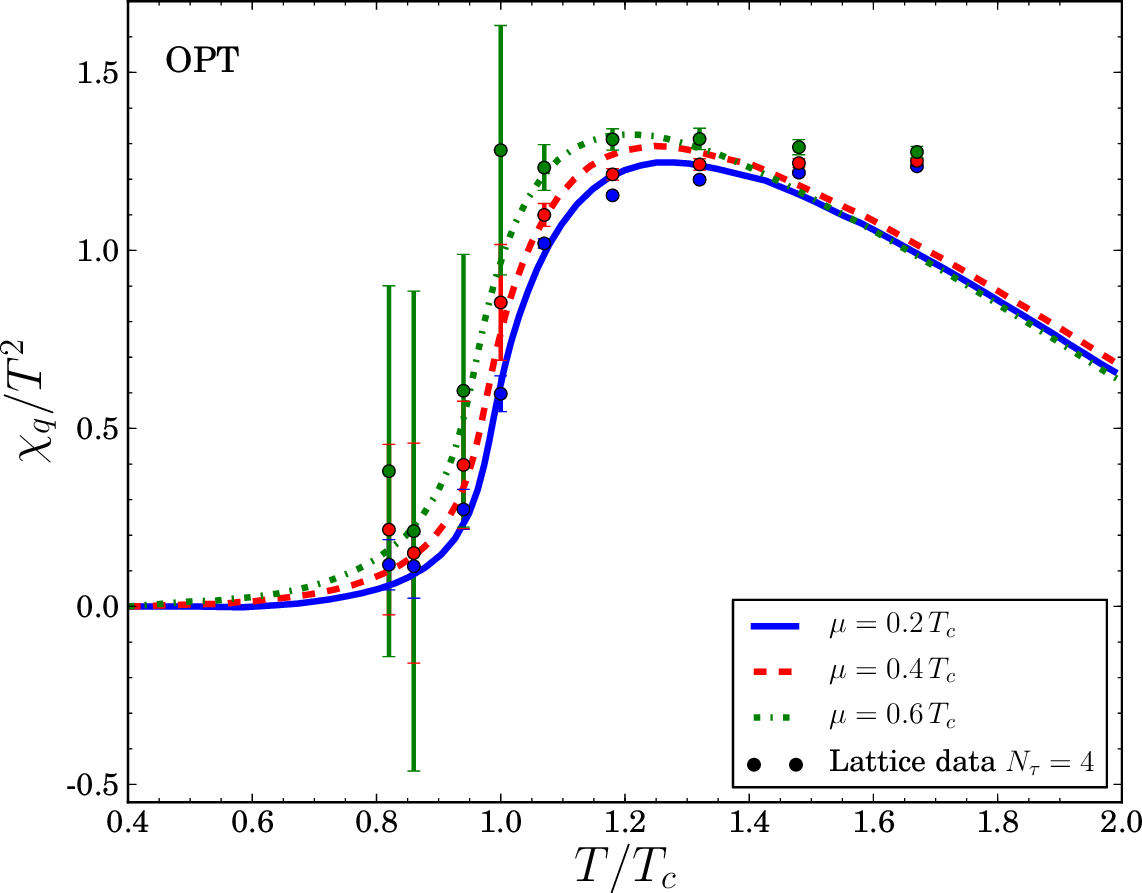} \quad \ 
 \end{subfigure}
 \caption{Normalized quark number susceptibility as a  function of $T/T_c$ for different values of $\mu$ as predicted by the LN approximation (top left panel), the 
 LNGv approximation  (top right panel) and the OPT (bottom panel). The model parameters are for $m_\pi=500\,{\rm MeV}$ and the lattice data were taken from Ref. \cite{ejiri}.}
  \label{c2PNJL}
\end{figure}

  \begin{figure}[h!]
 \begin{subfigure}
 \centering
 \includegraphics[width=.4\linewidth]{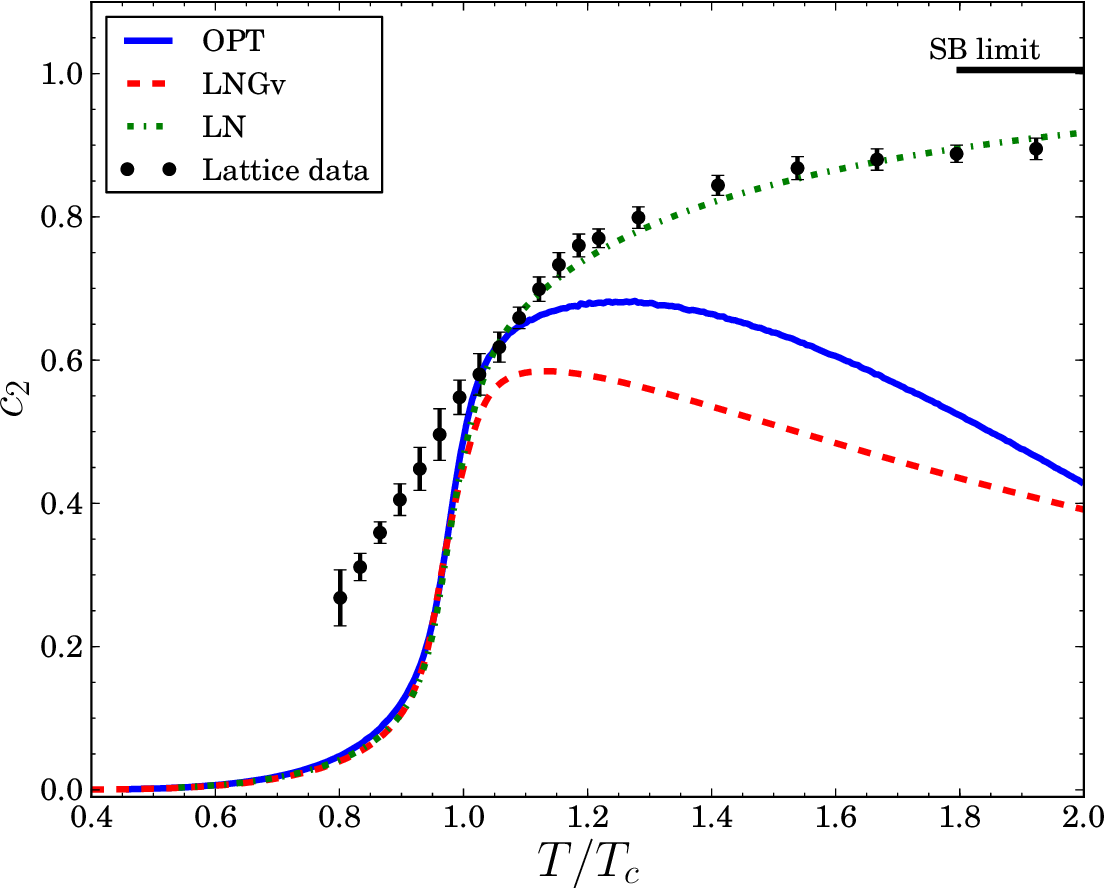} \quad \ 
 \end{subfigure}
 \begin{subfigure}
 \centering
 \includegraphics[width=.4\linewidth]{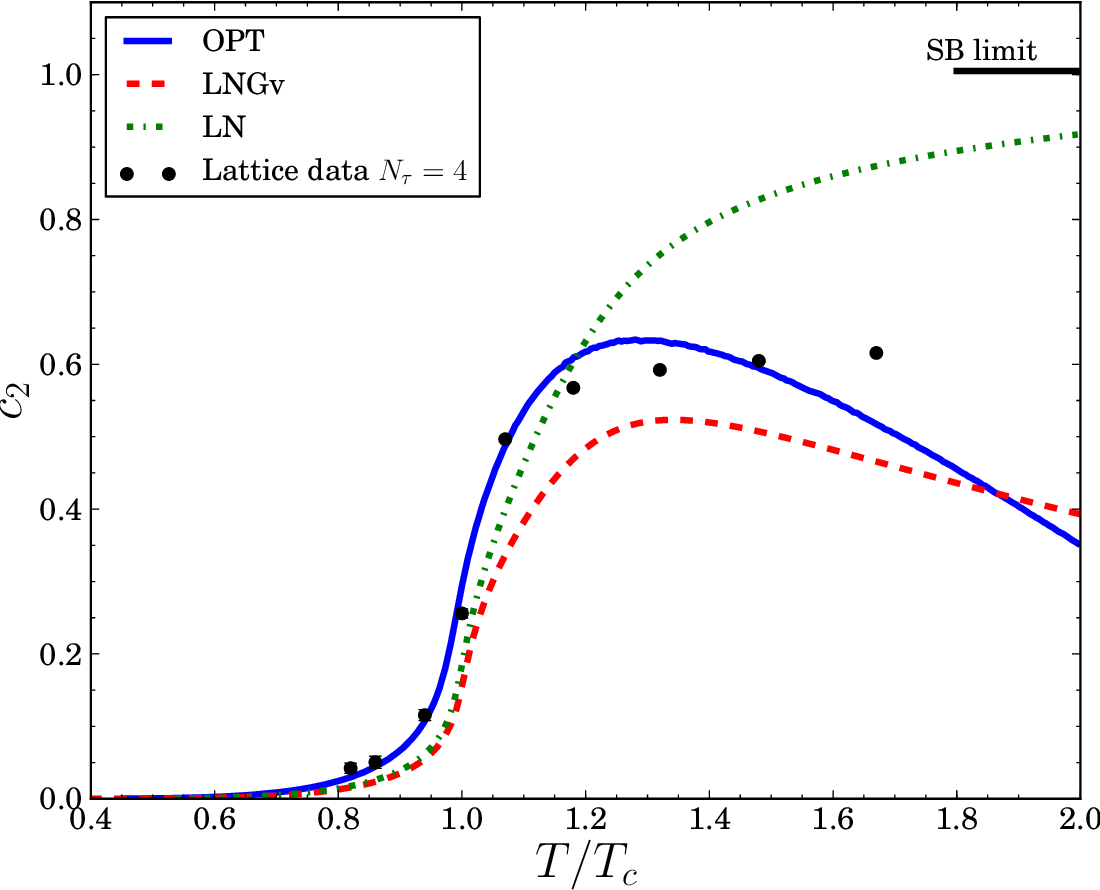} \quad \ 
 \end{subfigure}

 \caption{Taylor expansion coefficient, $c_2$, at $\mu=0$ as a function of $T/T_c$, obtained with the OPT and with the LN approximation with 
 $G_V=0$ and the LNGv approximation with $G_V=G_S/N_c$ for the PNJL model. Left panel: model parameters for $m_\pi=135\,{\rm MeV}$  and LQCD data taken from Ref. \cite{latticecont}. Right panel: model parameters for $m_\pi=500\,{\rm MeV}$  and LQCD data taken from Ref. \cite{ejiri}.}

\end{figure} 
   
\begin{figure}[h!]
  \centering
 \includegraphics[width=.4\linewidth]{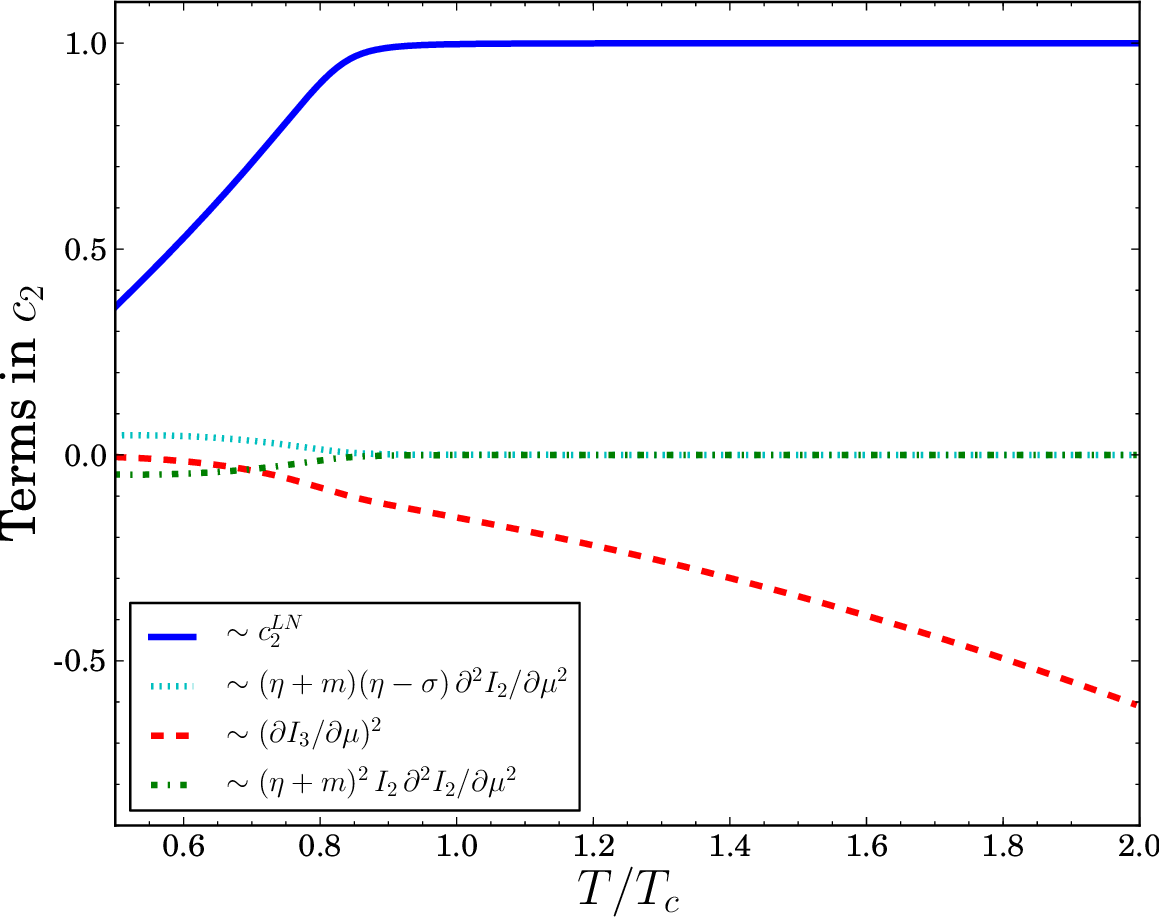} \quad \ 
 \caption{Terms contributing to $c_2$ as functions of $T/T_c$.}
   \label{densityOPT}
   \end{figure} 

 \subsection{ Temperature dependent couplings: the  EPNJL model.}
 
So far the couplings $G_S$ and $G_V$  were taken at constant values but, in order to mimic asymptotic freedom, they should decrease as energy scales, such as the temperature, rise. A way of implementing this behavior was advanced in Refs. \cite {epnjl1,epnjl2}, where $G_S$ was taken to be an effective vertex, $G_S (\Phi)$, which depends on $\Phi$. This new coupling is called the entanglement vertex, and the interactions are referred to as the entanglement interactions while the PNJL model plus the entanglement vertex is known as entangled PNJL (EPNJL) model.
A possible ansatz for $G_S(\Phi)$, and for $G_V(\Phi)$, is given by \cite {epnjl1,epnjl2}
\begin{equation}
G_S(\Phi) =G_S [ 1 - \alpha_1 \Phi {\bar \Phi} - \alpha_2 (\Phi^3 + {\bar \Phi}^3) ] \;\;,
\end{equation}
and
\begin{equation}
G_V(\Phi) =G_V [ 1 - \alpha_1 \Phi {\bar \Phi} - \alpha_2 (\Phi^3 + {\bar \Phi}^3) ] \;\;,
\end{equation}
which preserves chiral symmetry, C symmetry, and extended Z(3) symmetry. The parameters $\alpha_1$ and $\alpha_2$ are fixed in order to reproduce the
LQCD data which, at $\mu = 0$, show a coincidence between the pseudocritical temperatures for the chiral and confinement transitions. Here, following Refs. \cite {sugano,epnjl1} we adopt the values $\alpha_1 = \alpha_2 = 0.2$ together with the LNGv parametrization  given in table I. 
\begin{figure}[h!]
 \begin{subfigure}
 \centering
 \includegraphics[width=.4\linewidth]{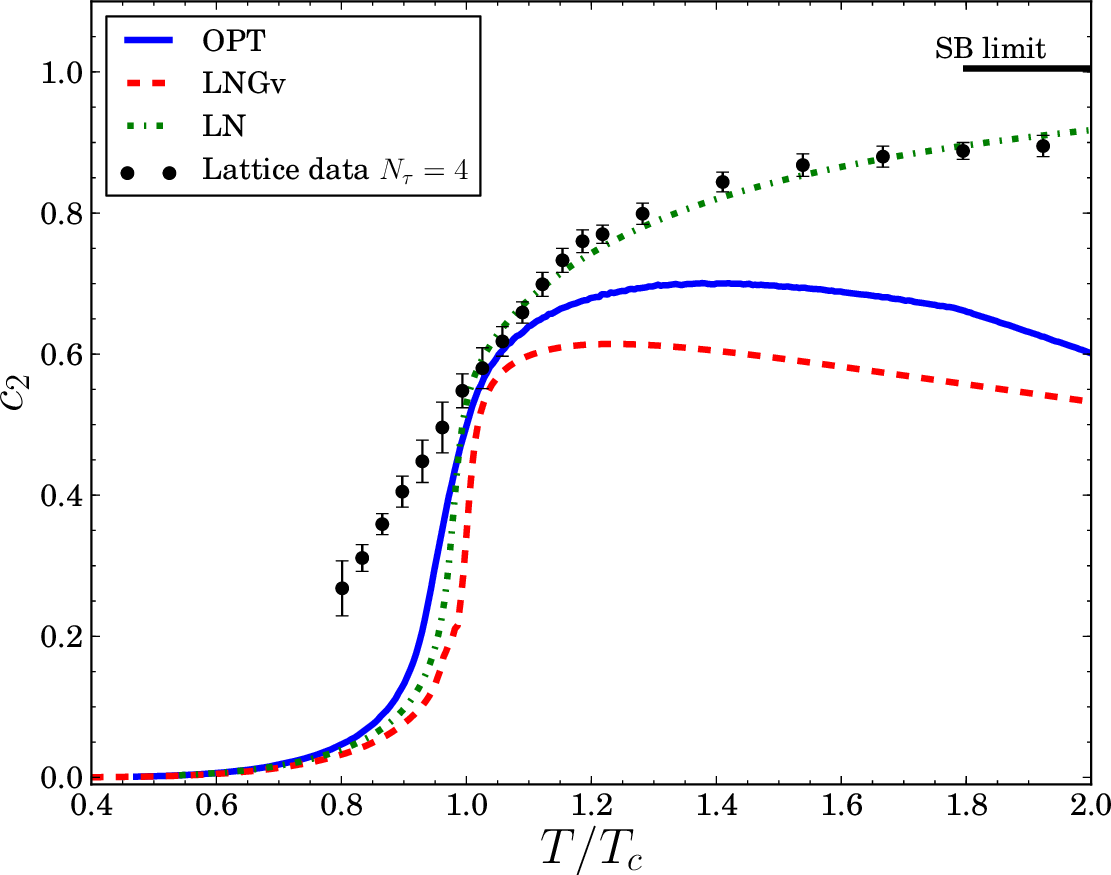} \quad \ 
 \end{subfigure}  
 \begin{subfigure}
 \centering
 \includegraphics[width=.4\linewidth]{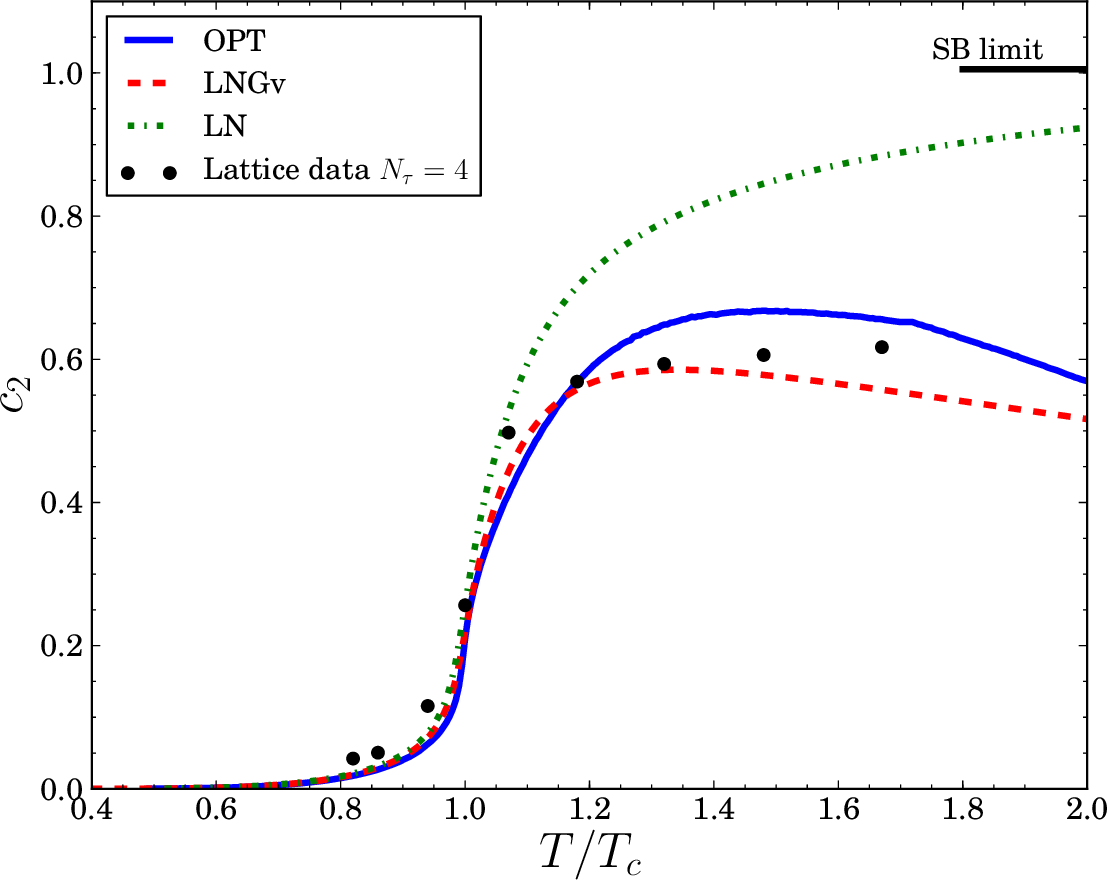} \quad \ 
 \end{subfigure} 
\caption{Taylor expansion coefficient, $c_2$, at $\mu=0$ as a function of $T/T_c$, obtained with the OPT and with the LN approximation with 
 $G_V=0$ and the LNGv approximation with $G_V=G_S/N_c$ for the EPNJL model. Left panel: model parameters for $m_\pi=135\,{\rm MeV}$  and LQCD data taken from Ref. \cite{latticecont}. Right panel: model parameters for $m_\pi=500\,{\rm MeV}$  and LQCD data taken from Ref. \cite{ejiri}.}
 
\end{figure}

Except for the replacements $G_S \to  G_S(\Phi)$ and $G_V \to G_V(\Phi)$ the OPT and the LN equations for the PNJL and for the EPNJL models are identical. Figure 6 shows the coefficient $c_2$ obtained with the OPT, the LNGv and the LN approximations for the EPNJL model. For the LNGv approximation we reproduce the behavior reported in a recent work by Sugano et al. \cite {sugano}. From a quantitative point of view, it seems that the results for the EPNJL model obtained with the OPT and the LNGv for $T < T_c$ are not so good as those obtained in the PNJL case. However, as stated in Ref. \cite {sugano} we are now interested in investigating if the $c_2$ maximum, which appears in the high-$T$ domain ($T > 1.2 \, T_c$), may be at least attenuated by the EPNJL coupling constant which decreases with the temperature. In this case, since maximum persists,  the problem is not completely solved. However, as expected, it is  now less pronounced and therefore give further support to the idea that  thermal effects on the PNJL couplings become to play an important role at high temperatures. Note that within this approach there are restrictions regarding the possible values of $\alpha_1$ and $\alpha_2$ \cite {epnjl1,epnjl2} so that with the values adopted here one may expect a reduction of about  $60 \%$ on the values of $G_S$ and $G_V$ above $T_c$. 

\section{Conclusions}

In this work the nonperturbative OPT method has been applied   to the PNJL model  in order to evaluate the pressure of hot and dense quark matter beyond the large-$N_c$ limit. The development of this type of alternative technique is important partially because the LQCD sign problem has not yet been fully circumvented and  partially because the LN approximation can furnish non accurate results at finite temperatures and/or densities \cite{bec, prdgn3d}.  As far as effective quark models are  concerned this method has been successful in describing chiral symmetry but it had never been used before in models which also display confinement. To perform such an application was one of our major goals. Another important remark is that, so far, this approximation has been used  mainly to draw the boundaries of QCD  phase diagram as well as to evaluate thermodynamical quantities without contrasting the results with available LQCD data as we have done here. When one considers the standard  PNJL Lagrangian  within the lagre-$N_c$ limit the pressure is written in terms of (scalar) quark condensates only so that important finite density effects are missed. Nevertheless, these  effects can still be incorporated within the same approximation at the expense of explicitly introducing a new vector channel with strength $G_V$ whose actual numerical value remains under dispute. Our work shows that finite $N_c$ radiative corrections may naturally generate a repulsive type of term with strength $G_S/(N_f N_c) $ as one could expect from the Fierz transformation properties \cite {klevansky}.  We have then evaluated the quark number density, $\rho_q$,  the quark number susceptibility, $\chi_q$, as well as the coefficient $c_2$ which appears when the quark pressure is represented by a Taylor series in powers of $\mu/T$. By comparing our results with LQCD predictions we have demonstrated how the OPT produces very accurate results up to about $1.2\,T_c$ for $\rho_q$, $\chi_q$ and $c_2$. Also, at higher temperatures, our predictions for $\rho_q$ remain superior to those furnished by the LN approximation especially as $\mu$ increases. However, for the quark number susceptibility $\chi_q$ (and the related $c_2$) the results produced by the approximations which contain vector repulsion start to deviate from the LQCD predictions at $T  \gtrsim 1.2 \,T_c$. 
We have identified  the mathematical origin of such a behavior  as being due to the presence of negative  contribution which is suppressed in the standard LN approximation. It is then possible to speculate that, at high temperatures, the apparent LN convergence towards the LQCD when $G_V=0$ could be accidental. In this case it is possible  that model approximation results for $c_2$, which include more finite $N_c$ effects, will oscillate around the LQCD predictions before convergence is achieved at higher orders (when vertex corrections  dress the couplings with thermal effects). Alternatively, the high-$T$ convergence towards the free gas result could be accelerated if the couplings present in the NJL type of  model were able to mimic the asymptotic freedom phenomenon observed in QCD.  
To back up this statement we recall that the Hard Thermal Loop Perturbation Theory, which is very similar to the OPT, has been recently applied  to QCD in order to evaluate quark susceptibilities up to three loops \cite {mike}. In this case, the numerical results show an excellent agreement with LQCD predictions as one would expect since the running of the coupling constant is naturally  taken into account within this particular application.  

 In order to explain the discrepancy  observed when applying the LN approximation to the PNJL model at $G_V\ne 0$ Schramm and Steinheimer concluded that one should expect a strong vector repulsion in the hadronic phase and near zero repulsion in the deconfined phase \cite {stefanGV2,stefanGV}.  Although our results basically support this hypothesis  we have explicitly shown here  that the same conclusion  will be reached even when $G_V=0$ since repulsive terms, parametrized by the scalar coupling $G_S$,  may also contribute to the pressure beyond the large-$N_c$ limit. In summary the presence of an explicit ($G_V \ne 0$) or a dynamically generated ($G_V=0$) repulsive contribution to the pressure is important for a realistic physical description within the confined phase. On the other hand our results suggest that within the deconfined phase the expected perturbative high-$T$ behavior can only be described by the PNJL type of effective model if the repulsive contribution, parametrized by $G_V$ or $G_S/N_c$, vanishes in that regime. It is possible that this will naturally happen when one considers higher order contributions such as vertex corrections. However, in practice this can turn out to be a hard exercise and then one can chose a more pragmatic alternative by requiring  the couplings to run with the temperature according to some ansatz  \cite {sugano}.

\acknowledgements
T.E.R., J.C.M and G.N.F. thank the Brazilian agencies CNPq and CAPES for MSc and PhD scholarships. M.B.P. is partially supported by CNPq and FAPESC. The authors thank Junpei Sugano and Jan Steinheimer for enlightening  discussions, and Shinji Ejiri for providing some of the lattice data. 

\appendix
\section{Color-trace over two loop contributions}\label{AppendixA}

In this Appendix we show how to calculate the color traces that appear in Eq. (\ref{320}). Let us start by denoting
\begin{align}
 S_{ii}&=\sum^\infty_{n=-\infty}\ln\left[\left(\omega_n-i\mu^\prime\right)^2+E_p^2\right]\notag\\
 &=\beta E_p+\ln\left[1+e^{-\beta\left(E_p-\mu^\prime\right)}\right]+\ln\left[1+e^{-\beta\left(E_p+\mu^\prime\right)}\right],
\end{align}
so that $S$ is a diagonal matrix.\par
Replacing $\mu^\prime=\mu-iA_{ii}$ and $\exp(iA_{ii})=L_{ii}$ we get
\begin{align}
 S_{ii}=\beta E_p+\ln\left[1+L_{ii}e^{-\beta\left(E_p-\mu\right)}\right]+\ln\left[1+L^*_{ii}e^{-\beta\left(E_p+\mu\right)}\right],
\end{align}
 where $A_{ii}$ is the $i$-th component of the diagonal matrix $A_4$ and $L_{ii}$
is the $i$-th component of the matrix $L$, which can be written in the diagonal form
\begin{align}
 L=\left[\begin{array}{c c c}
 e^{i\theta_1} & 0 & 0\\
 0 & e^{i\theta_2} & 0\\
 0 & 0 & e^{-i\left(\theta_1+\theta_2\right)}
          ×
         \end{array}\right].
\end{align}\par
Using the identity $\mathrm{Tr_c}\ln=\ln\det$ we get the the trace of the third term of Eq. (\ref{320}) which represents a one loop contribution
\begin{align}
 \mathrm{Tr_c}S=&N_c\beta E_p+\ln\left[g^+_l\left(E_p\right)\right]+\ln\left[g^-_l\left(E_p\right)\right],
\end{align}
where
\begin{align}
  g^+_l\left(E_p\right)&=1+3l e^{-\beta\left(E_p-\mu\right)}+3l^* e^{-2\beta\left(E_p-\mu\right)}+  e^{-3\beta\left(E_p-\mu\right)},\\
  g^-_l\left(E_p\right)&=1+3l^* e^{-\beta\left(E_p+\mu\right)}+3l e^{-2\beta\left(E_p+\mu\right)} +e^{-3\beta\left(E_p+\mu\right)}.
\end{align}
The next one loop term is 
\begin{align}
  \sum^\infty_{n=-\infty}\frac{1}{\left(\omega_n-i\mu^\prime\right)^2+ E_p^2}&=\frac{1}{2E_p×}\frac{\partial S_{ii}}{\partial E_p×}.
\end{align}
Therefore,
\begin{align}
 \mathrm{Tr_c}\sum^\infty_{n=-\infty}\frac{1}{\left(\omega_n-i\mu^\prime\right)^2+ E_p^2}&=\frac{1}{2E_p×}
 \frac{\partial \mathrm{Tr_c}S}{\partial E_p×}\notag\\
 &=\frac{N_c\beta}{2E_p×}\left[1-f^+_{l}-f^-_{l}\right],
\end{align}
where the Fermi distributions are given by
\begin{align}
 f^+_{l}\left(E_p\right)&=\frac{l e^{-\beta\left(E_p-\mu\right)}+2l^* e^{-2\beta\left(E_p-\mu\right)}+e^{-3\beta\left(E_p-\mu\right)}}
 { g^+_l\left(E_p\right)×},
 \end{align}
 and
 \begin{align}
  f^-_{l}\left(E_p\right)&=\frac{l^* e^{-\beta\left(E_p+\mu\right)}+2l e^{-2\beta\left(E_p+\mu\right)} +e^{-3\beta\left(E_p+\mu\right)}}
  {g^-_l\left(E_p\right)×}.  
\end{align}
The evaluation of the two loop contributions is more cumbersome as we show next. The fifth term of Eq. (\ref{320}) can be written as 
\begin{align}
 \sum^\infty_{n=-\infty}\frac{\omega_n-i\mu^\prime}{\left(\omega_n-i\mu^\prime\right)^2+E_p^2}
 \sum^\infty_{m=-\infty}\frac{\omega_m-i\mu^\prime}{\left(\omega_m-i\mu^\prime\right)^2+E_q^2}=\frac{i}{2×}
 \frac{\partial S_{ii}\left(E_p\right)}{\partial \mu^\prime×}
 \frac{i}{2×}\frac{\partial S_{ii}\left(E_q\right)}{\partial \mu^\prime×},
\end{align}
and 
\begin{align}
 \frac{\partial S_{ii}\left(q\right)}{\partial \mu^\prime×}\frac{\partial S_{ii}\left(p\right)}{\partial \mu^\prime×}=&\beta^2\left[
 \frac{e^{-\beta\left(E_p-\mu^\prime\right)}}{1+e^{-\beta\left(E_p-\mu^\prime\right)}×}-\frac{e^{-\beta\left(E_p+\mu^\prime\right)}}
 {1+e^{-\beta\left(E_p+\mu^\prime\right)}×}\right]\notag\\
 &\times\left[
 \frac{e^{-\beta\left(E_q-\mu^\prime\right)}}{1+e^{-\beta\left(E_q-\mu^\prime\right)}×}-\frac{e^{-\beta\left(E_q+\mu^\prime\right)}}
 {1+e^{-\beta\left(E_q+\mu^\prime\right)}×}\right]\notag\\
 &=\beta^2\left[
 \frac{L_{ii}e^{-\beta\left(E_p-\mu\right)}}{1+L_{ii}e^{-\beta\left(E_p-\mu\right)}×}-\frac{L^{*}_{ii}e^{-\beta\left(E_p+\mu\right)}}
 {1+L^{*}_{ii}e^{-\beta\left(E_p+\mu\right)}×}\right]\notag\\
 &\times\left[
 \frac{L_{ii}e^{-\beta\left(E_q-\mu\right)}}{1+L_{ii}e^{-\beta\left(E_q-\mu\right)}×}-\frac{L^*_{ii}e^{-\beta\left(E_q+\mu\right)}}
 {1+L^*_{ii}e^{-\beta\left(E_q+\mu\right)}×}\right] \,\,,\notag
 \end{align}
 \begin{align}
 \frac{\partial S_{ii}\left(q\right)}{\partial \mu^\prime×}\frac{\partial S_{ii}\left(p\right)}{\partial \mu^\prime×}&=\beta^2\left\lbrace
 \frac{L^2_{ii}e^{-\beta\left(E_p-\mu\right)}e^{-\beta\left(E_q-\mu\right)}}{\left[1+L_{ii}e^{-\beta\left(E_p-\mu\right)}\right]
 \left[1+L_{ii}e^{-\beta\left(E_q-\mu\right)}\right]×}\right.\notag\\
  &-\frac{e^{-\beta\left(E_p-\mu\right)}e^{-\beta\left(E_q+\mu\right)}}
 {\left[1+L_{ii}e^{-\beta\left(E_p-\mu\right)}\right]\left[1+L^*_{ii}e^{-\beta\left(E_q+\mu\right)}\right]×}\notag\\
  &-\frac{e^{-\beta\left(E_q-\mu\right)}e^{-\beta\left(E_p+\mu\right)}}
 {\left[1+L_{ii}e^{-\beta\left(E_q-\mu\right)}\right]\left[1+L^*_{ii}e^{-\beta\left(E_p+\mu\right)}\right]×}\notag\\
 &+\left.
 \frac{{L^{*}}^2_{ii}e^{-\beta\left(E_p+\mu\right)}e^{-\beta\left(E_q+\mu\right)}}{\left[1+L^*_{ii}e^{-\beta\left(E_p+\mu\right)}\right]
 \left[1+L^*_{ii}e^{-\beta\left(E_q+\mu\right)}\right]×}\right\rbrace\label{C10}.
\end{align}
After a straightforward but tediously long calculation of each of the four terms appearing in Eq. (\ref{C10}) we obtain
\begin{align}
 \mathrm{Tr_c}&\sum^\infty_{n=-\infty}\frac{\omega_n-i\mu^\prime}{\left(\omega_n-i\mu^\prime\right)^2+E_p^2}
 \sum^\infty_{m=-\infty}\frac{\omega_m-i\mu^\prime}{\left(\omega_m-i\mu^\prime\right)^2+E_q^2}\notag\\
 &=-\frac{N_c\beta^2}{4×}\left\lbrace\left[f^+_{l}\left(E_p\right)-f^-_{l}\left(E_p\right)\right]\left[f^+_{l}\left(E_q\right)-f^-_{l}\left(E_q\right)\right]
 +\Delta_3^2\right\rbrace,
\end{align}
where
\begin{align}
 \Delta_3^2&=\frac{e^{-\beta\left(E_p-\mu\right)}e^{-\beta\left(E_q-\mu\right)}}{g^+_l(E_p)g^+_l(E_q)}\left\lbrace2\left(l^2-l^*\right)+\left(ll^*-1\right)
 \left[e^{-\beta\left(E_p-\mu\right)}+e^{-\beta\left(E_q-\mu\right)}\right]\right.\notag\\
 &+\left.2\left({l^*}^2-l\right)e^{-\beta\left(E_p-\mu\right)}e^{-\beta\left(E_q-\mu\right)}\right\rbrace\notag\\
 &-\frac{e^{-\beta\left(E_p-\mu\right)}e^{-\beta\left(E_q+\mu\right)}}{g^+_l(E_p)g^-_l(E_q)}\left\lbrace2\left(l-{l^*}^2\right)
 e^{-\beta\left(E_p-\mu\right)}+2\left(l^*-l^2\right) e^{-\beta\left(E_q+\mu\right)}\right.\notag\\
 &+\left.\left(1-ll^*\right)\left[1+e^{-\beta\left(E_p-\mu\right)}e^{-\beta\left(E_q+\mu\right)}
 \right]\right\rbrace\notag\\
 &-\frac{e^{-\beta\left(E_q-\mu\right)}e^{-\beta\left(E_p+\mu\right)}}{g^+_l(E_q)g^-_l(E_p)}\left\lbrace2\left(l-{l^*}^2\right)
 e^{-\beta\left(E_q-\mu\right)}+2\left(l^*-l^2\right) e^{-\beta\left(E_p+\mu\right)}\right.\notag\\
 &+\left.\left(1-ll^*\right)\left[1+e^{-\beta\left(E_q-\mu\right)}e^{-\beta\left(E_p+\mu\right)}
 \right]\right\rbrace\notag\\
 &+\frac{e^{-\beta\left(E_p+\mu\right)}e^{-\beta\left(E_q+\mu\right)}}{g^-_l(E_p)g^-_l(E_q)}\left\lbrace2\left({l^*}^2-l\right)+\left(ll^*-1\right)
 \left[e^{-\beta\left(E_p+\mu\right)}+e^{-\beta\left(E_q+\mu\right)}\right]\right.\notag\\
 &+\left.2\left(l^2-l^*\right)e^{-\beta\left(E_p+\mu\right)}e^{-\beta\left(E_q+\mu\right)}\right\rbrace.
\end{align}
Finally, for the last term of Eq. (\ref{320}) we have 
\begin{align}
 \sum^\infty_{n=-\infty}\frac{1}{\left(\omega_n-i\mu^\prime\right)^2+E_p^2}
 \sum^\infty_{m=-\infty}\frac{1}{\left(\omega_m-i\mu^\prime\right)^2+E_q^2}=\frac{1}{2E_p×}
 \frac{\partial S_{ii}\left(E_p\right)}{\partial E_p×}
 \frac{1}{2E_q×}\frac{\partial S_{ii}\left(E_q\right)}{\partial E_q×}.
\end{align}
Then,
\begin{align}
 \frac{\partial S_{ii}\left(E_p\right)}{\partial E_p×}\frac{\partial S_{ii}\left(E_q\right)}{\partial E_q×}=&\beta^2\left[1-
 \frac{e^{-\beta\left(E_p-\mu^\prime\right)}}{1+e^{-\beta\left(E_p-\mu^\prime\right)}×}-\frac{e^{-\beta\left(E_p+\mu^\prime\right)}}
 {1+e^{-\beta\left(E_p+\mu^\prime\right)}×}\right]\notag\\
 &\times\left[1-
 \frac{e^{-\beta\left(E_q-\mu^\prime\right)}}{1+e^{-\beta\left(E_q-\mu^\prime\right)}×}-\frac{e^{-\beta\left(E_q+\mu^\prime\right)}}
 {1+e^{-\beta\left(E_q+\mu^\prime\right)}×}\right]\notag\\
 &=\beta^2\left[1-
 \frac{L_{ii}e^{-\beta\left(E_p-\mu\right)}}{1+L_{ii}e^{-\beta\left(E_p-\mu\right)}×}-\frac{L^{*}_{ii}e^{-\beta\left(E_p+\mu\right)}}
 {1+L^{*}_{ii}e^{-\beta\left(E_p+\mu\right)}×}\right]\notag\\
 &\times\left[1-
 \frac{L_{ii}e^{-\beta\left(E_q-\mu\right)}}{1+L_{ii}e^{-\beta\left(E_q-\mu\right)}×}-\frac{L^*_{ii}e^{-\beta\left(E_q+\mu\right)}}
 {1+L^*_{ii}e^{-\beta\left(E_q+\mu\right)}×}\right].
\end{align}
Finally, the remaining trace gives
\begin{align}
 \mathrm{Tr_c}&\sum^\infty_{n=-\infty}\frac{1}{\left(\omega_n-i\mu^\prime\right)^2+E_p^2}
 \sum^\infty_{m=-\infty}\frac{1}{\left(\omega_m-i\mu^\prime\right)^2+E_q^2}\notag\\
 &=\frac{N_c\beta^2}{4E_pE_q×}\left\lbrace\left[1-f^+_{l}\left(E_p\right)-f^-_{l}\left(E_p\right)\right]\left[1-f^+_{l}\left(E_q\right)-f^-_{l}\left(E_q\right)\right]
 +\Delta_2^2\right\rbrace,
\end{align}
where
\begin{align}
 \Delta_2^2&=\frac{e^{-\beta\left(E_p-\mu\right)}e^{-\beta\left(E_q-\mu\right)}}{g^+_l(E_p)g^+_l(E_q)}\left\lbrace2\left(l^2-l^*\right)+\left(ll^*-1\right)
 \left[e^{-\beta\left(E_p-\mu\right)}+e^{-\beta\left(E_q-\mu\right)}\right]\right.\notag\\
 &+\left.2\left({l^*}^2-l\right)e^{-\beta\left(E_p-\mu\right)}e^{-\beta\left(E_q-\mu\right)}\right\rbrace\notag\\
 &+\frac{e^{-\beta\left(E_p-\mu\right)}e^{-\beta\left(E_q+\mu\right)}}{g^+_l(E_p)g^-_l(E_q)}\left\lbrace2\left(l-{l^*}^2\right)
 e^{-\beta\left(E_p-\mu\right)}+2\left(l^*-l^2\right) e^{-\beta\left(E_q+\mu\right)}\right.\notag\\
 &+\left.\left(1-ll^*\right)\left[1+e^{-\beta\left(E_p-\mu\right)}e^{-\beta\left(E_q+\mu\right)}
 \right]\right\rbrace\notag\\
 &+\frac{e^{-\beta\left(E_q-\mu\right)}e^{-\beta\left(E_p+\mu\right)}}{g^+_l(E_q)g^-_l(E_p)}\left\lbrace2\left(l-{l^*}^2\right)
 e^{-\beta\left(E_q-\mu\right)}+2\left(l^*-l^2\right) e^{-\beta\left(E_p+\mu\right)}\right.\notag\\
 &+\left.\left(1-ll^*\right)\left[1+e^{-\beta\left(E_q-\mu\right)}e^{-\beta\left(E_p+\mu\right)}
 \right]\right\rbrace\notag\\
 &+\frac{e^{-\beta\left(E_p+\mu\right)}e^{-\beta\left(E_q+\mu\right)}}{g^-_l(E_p)g^-_l(E_q)}\left\lbrace2\left({l^*}^2-l\right)+\left(ll^*-1\right)
 \left[e^{-\beta\left(E_p+\mu\right)}+e^{-\beta\left(E_q+\mu\right)}\right]\right.\notag\\
 &+\left.2\left(l^2-l^*\right)e^{-\beta\left(E_p+\mu\right)}e^{-\beta\left(E_q+\mu\right)}\right\rbrace.
\end{align}

\end{document}